\documentclass[preprint,12pt]{elsarticle}
\pdfoutput=1

\usepackage[T1]{fontenc} 
\usepackage[utf8]{inputenc}
\usepackage{color}
\usepackage{amsmath,amssymb}
\usepackage{breqn}
\usepackage{graphicx}
\usepackage{float}
\usepackage{hyperref}
\usepackage[normalem]{ulem}
\usepackage{soul,xcolor}
\usepackage{rotating}
\usepackage{array,multirow}
\usepackage{geometry}
\journal{Nuclear Physics B}

\begin{document}

\begin{frontmatter}



\title{Leptogenesis and low-energy CP violation in a type-II-dominated
left-right seesaw model}

\author[inst1]{Thomas~Rink\fnref{label2}}
\fntext[label2]{Corresponding author}
\ead{thomas.rink@mpi-hd.mpg.de}

\author[inst1]{Werner~Rodejohann}
\ead{werner.rodejohann@mpi-hd.mpg.de}

\affiliation[inst1]{organization={Max-Planck-Institut f\"ur Kernphysik},
            addressline={Saupfercheckweg 1}, 
            city={Heidelberg},
            postcode={69117},
            country={Germany}}

\author[inst2]{Kai~Schmitz}
\ead{kai.schmitz@cern.ch}

\affiliation[inst2]{organization={Theoretical Physics Department},
            addressline={CERN}, 
            city={1211 Geneva 23},
            country={Switzerland}}

\begin{abstract}
We consider leptogenesis in a left-right-symmetric seesaw scenario in which neutrino mass generation and leptogenesis are dominated by the type-II seesaw term.
Motivated by grand unification, we assume that the neutrino Dirac mass matrix is dominated by a single entry of the order of the top-quark mass, which leaves the low-energy phases of the lepton mixing matrix as the only sources of CP violation. 
Working in a regime where the triplet scalar predominantly decays into leptons, this results in a predictive scenario based on a minimal number of parameters.
We perform a detailed analysis of the flavored Boltzmann equations within a revised density matrix framework and 
demonstrate that the observed baryon asymmetry can be successfully generated in this simple model.
We point out that the significance of flavor effects is limited, and we discuss the implications for low-energy observables such as the Dirac CP phase and neutrinoless double beta decay.
\end{abstract}


\end{frontmatter}



\section{Introduction}
\label{sec:introduction}


\noindent The explanation of the observed baryon-to-photon ratio is one of the most fascinating and important questions of fundamental physics since it is directly linked to the origin of our existence.
The underlying conditions for successful baryogenesis formulated by Sakharov~\cite{Sakharov:1967dj} allow for a wide variety of models, among which leptogenesis~\cite{Fukugita:1986hr} is of special interest because it establishes a connection between the baryon asymmetry of the Universe (BAU) and the generation of light active neutrino masses.
In leptogenesis scenarios, a particle--antiparticle asymmetry is first created in the lepton sector by CP-violating decays of heavy states at temperatures around their mass, before it is transferred to the baryon sector by Standard Model (SM) sphaleron processes~\cite{Klinkhamer:1984di,Arnold:1987mh,Arnold:1987zg}. 
In its standard formulation, leptogenesis proceeds via the decay of right-handed neutrinos, the same particles that are at the origin of neutrino mass in the type-I seesaw mechanism~\cite{Minkowski:1977sc,Yanagida:1979as,Yanagida:1980xy,GellMann:1980vs,Mohapatra:1980yp}.
Remarkably enough, a large number of neutrino mass models in the literature allows for successful 
leptogenesis~\cite{Hambye:2012fh}, which assigns neutrinos a special role in the cosmological history of our Universe.
Moreover, as the option of measuring CP violation in the lepton sector becomes more and more realistic, and first hints towards nontrivial values of the CP phase $\delta$ have emerged~\cite{Abe:2019vii, Capozzi:2020qhw}, the question whether there is a connection between low-energy and high-energy CP violation is imminent. 
For these reasons, leptogenesis has become a very active field in the last decades, and many phenomenological as well as formal aspects have been investigated in great detail. 
For more insights, see the review articles in Refs.~\cite{Buchmuller:2004nz,Davidson:2008bu,Fong:2013wr,Chun:2017spz} and references therein.
%
%
Neutrino masses may also be generated by mechanisms beyond the type-I seesaw, in particular, the type-II or -III seesaw, which  include scalar~\cite{Mohapatra:1980yp,Magg:1980ut, Schechter:1980gr, Wetterich:1981bx, Lazarides:1980nt, Kanemura:2012rs} or fermionic \cite{Foot:1988aq, Ma:1998dn, Ma:2002pf} triplets, respectively. 
In these mechanisms, the BAU can be generated as well. Indeed, leptogenesis through decays of scalar~\cite{ODonnell:1993obr,Ma:1998dx,Hambye:2003ka,Antusch:2004xy,Lavignac:2015gpa} and fermionic~\cite{Hambye:2003rt,Fischler:2008xm,Strumia:2008cf} triplets has been discussed. 
Mixed models have also been investigated~\cite{Joshipura:1999is,Joshipura:2001ya,Rodejohann:2004cg,Hallgren:2007nq,Abada:2008gs,Akhmedov:2008tb,Borah:2013bza,Parida:2016hln, Cogollo:2019mbd,Ferreira:2019qpf,Chakraborty:2019uxk}.
In this case, both neutrino mass and lepton asymmetry receive more than one contribution.


Our work deals with a minimal realization of such a mixed model, where the neutrino mass arises from a combined type-I and type-II seesaw.
Note that one scalar triplet alone is not sufficient to generate a CP asymmetry, as there needs to be an additional particle to generate a loop diagram with which the tree-level triplet decay can interfere.
In our case, the right-handed neutrinos of the type-I seesaw contribution play this role. 
Leptogenesis within our approach is thus generically governed by both seesaw contributions and their underlying couplings, namely, the Dirac Yukawa couplings from the type-I sector as well as the triplet Yukawa and trilinear scalar couplings from the type-II sector. 
Working in the parameter region where the type-II contribution dominates leptogenesis and neutrino mass, the diagram that generates the lepton asymmetry is the decay of the triplet with a vertex correction involving right-handed neutrinos. 
We then make the simplifying assumption that there is "left-right seesaw symmetry"~\cite{Joshipura:2001ya,Joshipura:2001ui,Akhmedov:2005np,Akhmedov:2006de}, such that the right-handed neutrino mass matrix is proportional to the triplet Yukawa matrix. 
Within type-II dominance, this renders the light- and heavy-neutrino mass matrices proportional to each other, thereby fixing the heavy-flavor sector up to a normalization. 
As a consequence, the Pontecorvo--Maki--Nakagawa--Sakata (PMNS) lepton mixing matrix diagonalizes both the left- and right-handed Majorana mass matrices simultaneously.
What remains to be fixed is the Yukawa matrix that couples left- and right-handed fermions, i.e., the Dirac mass matrix of the type-I seesaw term. 
Assuming a relation to the up-quark sector as it occurs in grand unified theories (GUTs) leads it to be dominated by one single entry, the top-quark mass. 
Thus, in what regards the flavor parameters, only the elements of the PMNS matrix, in particular, the CP phases influence the final baryon asymmetry. 
The special feature of the model under investigation is a very small number of free parameters and its direct link between the baryon asymmetry and measured, or measurable, low-energy neutrino observables.

In this paper, we discuss the requirements underlying a type-II-dominated light neutrino mass and perform a detailed analysis of the relevant Boltzmann equations in the density matrix formalism. 
We work in a regime of parameter space where flavor effects are only of minor importance. This, in turn, allows for a careful and robust analysis of the dependence of the baryon asymmetry on measurable low-energy parameters, in particular, the mass ordering, the CP phase $\delta$, and the effective neutrino mass $m_{ee}$ that is crucial for neutrinoless double beta decay. 
Additional parameters are the mass of the lightest neutrino and the mass of the scalar triplet. 
The requirement of type-II seesaw dominance provides additional constraints on these quantities. \\


\noindent This paper is structured as follows:
Sec.~\ref{sec:seesaws} summarizes the generation of neutrino mass in the minimal mixed type-(I$+$II) seesaw model and the simplifications brought about by a dominating triplet contribution. 
The corresponding leptogenesis scenario is covered in Sec.~\ref{sec:leptogenesis}, both analytically and numerically. 
We present our results in Sec.~\ref{sec:results} and conclude with Sec.~\ref{sec:conclusions}.
Various technical details are collected in a number of appendices.


\section{Minimal mixed neutrino mass model with type-II dominance}
\label{sec:seesaws}


\noindent The model under consideration generates light neutrino masses through Yukawa interactions with right-handed (RH) heavy neutrinos possessing a Majorana mass term and through Yukawa interactions with a scalar isospin triplet.
Consequently, both type-I and type-II seesaw contributions to the light neutrino mass matrix $m_\nu$ are present:
\begin{equation}
m_\nu = m_{II} + m_I = m_{II} - m_D^{T} \, m_N^{-1} \, m_D \,.
\end{equation}
In the charged-lepton basis, $m_\nu$ is diagonalized by the PMNS matrix $U$: 
\begin{align}
\label{eq:PMNS}
U&=\begin{pmatrix}
 c_{12}c_{13} & s_{12}c_{13} & s_{13}e^{-i \delta} \\
 -s_{12}c_{23} -c_{12}s_{13}s_{23} e^{i \delta} & c_{12}c_{23} - 
 s_{12}s_{13}s_{23} e^{i \delta} 
 & c_{13}s_{23} \\
 s_{12}s_{23} -c_{12}s_{13}c_{23} e^{i \delta} & -c_{12}s_{23} -
 s_{12}s_{13}c_{23} e^{i \delta}
 & c_{13}c_{23} 
\end{pmatrix}
\times \text{diag}(1, e^{i \sigma},e^{i \tau})\, ,
\end{align}
with $c_{ij} = \cos \theta_{ij}$ and $s_{ij} = \sin  \theta_{ij}$.
In what follows, we will use the recent fit results from Ref.~\cite{Capozzi:2020qhw} for the mixing angles $\theta_{ij}$, the CP phase $\delta$, and the mass-squared differences $\delta m^{2}=m_{2}^{2}-m_{1}^{2}$, $\Delta m^{2}=m_{3}^{2}-(m_{2}^{2}+m_{1}^{2})/2$, which determine the individual neutrino masses once the smallest mass $m_{\rm sm}$ is fixed ($m_{\rm sm} = m_1$ for the normal mass ordering, $m_{\rm sm} = m_3$ for the inverted one).
The absolute neutrino mass scale is bounded from above by cosmological observations to sub-eV values, with the precise upper bound depending on the data sets and cosmological assumptions~\cite{Aghanim:2018eyx}.
The Majorana phases $\sigma$ and $\tau$ are unconstrained.
We will assume type-II dominance in $m_\nu$ in this paper.
That is, we will assume that the main contribution to light neutrino masses comes from interactions with the scalar triplet, 
\begin{align}
    m_\nu \simeq m_{II}\,.
\end{align}
Hence, the PMNS matrix $U$ diagonalizes $m_{II}$, which fixes the relevant triplet Yukawa matrix; see Eq.~\eqref{eq:LRnumasses} below. 
With the additional assumption of a discrete left-right (LR) symmetry, $m_{II} \propto m_N$, the flavor structure of the heavy RH neutrino mass matrix is fixed.
The remaining unknown flavor structure comes from the Dirac mass matrix of the type-I seesaw contribution, which will be assumed to be strongly hierarchical.
Overall, the situation is then quite minimal and predictive.
Let us now discuss our framework in more detail.


\subsection{Mixed type-II-dominated seesaw model}
\label{sec:MixedTypeIIDom}


\noindent Type II-dominance within our combined seesaw framework generically implies the type-I scale to be higher than the type-II one, $\Lambda_{\rm I} \gg \Lambda_{\rm II}$.
The assumption that the scalar triplet (type-II) contribution to neutrino mass is proportional to the right-handed neutrino mass matrix, $m_{II} \propto m_N$~\cite{Joshipura:2001ya,Joshipura:2001ui,Akhmedov:2005np,Akhmedov:2006de} could originate, for instance, in left-right-symmetric theories based on $SU(2)_L \times SU(2)_R \times U(1)_{B-L}$, in which $m_N$ is given by $2 \, k \, v_R$, where $v_R$ is the VEV of a scalar $SU(2)_R$ triplet and $k$ a Yukawa matrix.
The discrete left-right symmetry in such models then leads to $f = k$, where $f$ is the Yukawa matrix of the $SU(2)_L$ triplet, cf.\ Eq.~\eqref{eq:typeIILagra}.
Although we remain agnostic about the explicit realization of such a framework, this assumption implies useful relations that help us shrink the parameter space down to a few variables, most of which are measurable at low energies.
Our calculation is assuming that the underlying sector that leads to the discrete left-right symmetry has no impact on leptogenesis.
Within a left-right-symmetric model, it could easily arise when the gauge bosons and right-handed triplets are much heavier than the RH neutrino masses and the $SU(2)_L$ triplets. 
Flavor symmetry approaches that could lead to our scenario can also be constructed along the line of Ref.~\cite{Xing:2019edp}.
We leave an investigation of an explicit realization as well as a study of leptogenesis in other areas of parameter space (i.e., when diagrams other than triplet decay dominate leptogenesis) for future work.


\paragraph{Type-I seesaw contribution}

Within our model, light neutrino masses are partially generated by the interplay of neutrino Dirac mass terms, induced by Yukawa interactions of left-handed (LH) and RH neutrinos, and Majorana mass terms of the RH singlet neutrinos. 
The corresponding type-I Lagrangian is given by 
\begin{align}
\mathcal{L}_{\text{I}} \supset -\frac{1}{2} \left( m_N \right)_{AB} (\overline{N_R})_{A}\, \mathcal{C}^\dagger\, (\overline{N_R}^{T})_{B} - h_{Aj} \left(\overline{N_{R}}\right)_{A} \widetilde{H}^{\dagger} L_{j}  + \textrm{h.c.},
\end{align}
where $\mathcal{C}$ represents the charge conjugation matrix, and with the SM Higgs doublet $H = \left(H^{+},H^{0}\right)^{T}$,  its conjugate $\widetilde{H}=i\sigma_{2} H^{*}$, the RH neutrino singlets $N_R$ and the LH lepton doublets $L$.
Combining Dirac and Majorana masses into one mass matrix and diagonalizing 
it leads to the well-known type-I seesaw mass formula  
\begin{align}\label{eq:typeImass}
	m_{I}\simeq -m_{D}^{T}\ m_{N}^{-1} m_{D}=-M_{D}^{T}\ D_{N}^{-1}\ M_{D}\, ,
\end{align}
with $M_{D}=U^{T}\,m_{D}$ and $D_{N}^{-1}=U^{\dagger}m_{N}^{-1}\,U^{*}$.
The masses of the right-handed neutrinos will be denoted $M_{1,2,3}$, while the Dirac mass matrix in the RH neutrino mass basis is defined as $m_D = y\, v_{\rm ew}/\sqrt{2}$ with the Higgs vacuum expectation value (VEV) $v_{\rm ew} = \sqrt{2}\,\langle H^{0} \rangle = 246$ GeV and $y\simeq U^{\dagger}h$ being the corresponding Yukawa couplings.


\paragraph{Type-II seesaw contribution}

Since we assume type-II dominance, the corresponding seesaw contribution will be of main importance in the course of this work. 
It is given by massive scalar isospin triplets coupling to LH leptons and the SM Higgs doublet. 
For our purposes, we only need one massive scalar triplet such that the relevant part of the type-II Lagrangian is of the form~\cite{Hambye:2003rt, Hambye:2003ka, Sierra:2014tqa, Lavignac:2015gpa} 
\begin{align}\label{eq:typeIILagra}
\begin{aligned}
\mathcal{L}_{II}\supset & -\left( f_{\alpha\beta}\, L_{\alpha}^{T}\, \mathcal{C}\, i\sigma_{2}\, \Delta\, L_{\beta} + \mu\, H^{T} i\sigma_{2}\, \Delta^{\dagger}\, H + \textrm{h.c.}\right)
- M_{\Delta}^{2} \, {\rm Tr} \left[\Delta^{\dagger}\Delta\right] ,
\end{aligned}
\end{align}
where the triplet is given by
\begin{align}\label{eq:triplet}
\Delta=\begin{pmatrix} \Delta^{+}/\sqrt{2} & \Delta^{++}\\ 
\Delta^{0} & - \Delta^{+}/\sqrt{2} \end{pmatrix}.
\end{align}
Upon symmetry breaking, the neutral component $\Delta^0$ develops a nonzero VEV
\begin{equation}
\label{eq:vL}
v_L = \left<\Delta^0\right> = \frac{\mu\,v_{\rm ew}^2}{2\,M_\Delta^2} \,,
\end{equation}
which depends on the trilinear coupling $\mu$, the triplet mass $ M_{\Delta}$, and the Higgs VEV $v_{\rm ew}$, and which induces the following contribution to the light neutrino mass:
\begin{align}
\label{eq:typeIImass}
m_{II} = 2\, v_L f = \frac{\mu\,v_{\rm ew}^2}{M_\Delta^2} f \,.
\end{align}

\noindent For leptogenesis, it is important to note that the triplet has two different 
decay channels, a leptonic and a bosonic one, whose tree-level decay rates, for one triplet degree of freedom, are 
given by
\begin{align}
\label{eq:tripletDecayRates_lepton}
\Gamma\left(\Delta_{i}\rightarrow\overline{L}\overline{L}\right) & = \frac{1}{8\pi}
\lambda_{L}^{2}\,M_{\Delta}\,, & \lambda_{L} & \equiv \sqrt{{\rm Tr}(ff^{\dagger})}\,
,\\
\label{eq:tripletDecayRates_higgs}
\Gamma\left(\Delta_{i}\rightarrow HH \right) & = \frac{1}{8\pi} \lambda^{2}_{H}\,M_{\Delta}\,, & \lambda_{H} & \equiv\frac{|\mu|}{M_{\Delta}}\, .
\end{align}
Making use of the couplings $\lambda_L$ and $\lambda_H$, one obtains handy expressions for the corresponding branching ratios, $B_{L,H}=\lambda_{L,H}^{2}/(\lambda_{L}^{2}+\lambda_{H}^{2})$.


\paragraph{Connection between different seesaw sectors}

Due to the underlying LR symmetry, LH and RH neutrinos are subject to the same coupling matrix, 
\begin{align}
\label{eq:LRnumasses}
m_{N}\equiv 2\,  v_{R}\, f \,, \qquad m_{II}= 2\, v_{L} \, f \,,
\end{align}
where $v_R$ is defined here as the overall mass scale of $m_N$, and $v_L$ was defined in Eq.~\eqref{eq:vL}. 

\noindent Equations~\eqref{eq:LRnumasses} further imply
\begin{align}
m_{N}=\frac{m_{II}}{r}\simeq\frac{m_{\nu}}{r}\ , \quad\text{with } r\equiv\frac{v_{L}}{v_{R}}\,.
\end{align}
This simple relation illustrates that $m_N$ and $m_\nu$ can be diagonalized by the same matrix, which, in the charged-lepton basis, is exactly the PMNS matrix $U$ and we immediately see that both mass hierarchies are also identical,
\begin{equation}
\label{eq:linkedMasses}
D_{N} = \text{diag}\left(M_{1},M_{2},M_{3}\right) \simeq \frac{1}{r}\, \text{diag}\left(m_{1},m_{2},m_{3}\right) = \frac{1}{r}\, D_{\nu}\, ,
\end{equation}
upon using $D_{\nu,N} =  U^{T}m_{\nu,N}\,U$.
The remaining flavor structure resides in the Dirac mass matrix $m_D$.
It is natural to assume that it is related to the up-quark mass matrix.
This is what can be realized in $SO(10)$ models, which interestingly can be broken down to the SM via an intermediate $SU(2)_L \times SU(2)_R \times U(1)_{B-L}$ step, see Ref.~\cite{Deppisch:2017xhv} for a recent analysis. 
To be more precise, a $SO(10)$-like dominance of the $\mathbf{10}$ Higgs representation can be used to establish this connection between the Dirac neutrino and up-type quark mass matrices~\cite{Dutta:2004hp,Bertolini:2004eq}. 
We thus assume
\begin{align}
\label{eq:DiracNuMass}
m_{D} = c \cdot \text{diag}\left(m_{u},m_{c},m_{t}\right) 
\simeq c \cdot \text{diag}\left(0,0,m_{t}\right)\, ,
\end{align}
with the factor c being a quantity of $\mathcal{O}(1)$ that depends on the underlying explicit model building.
For instance, in a minimal SO(10) framework with \textbf{10} and \textbf{126} Higgs representations $c=3$ \cite{Georgi:1979df}.
However, since the resulting mixing angles are in conflict with the experimental measurements \cite{Harvey:1981hk}, additional model building is required.
At this point, we leave model building aside and assume a generic value of $c\sim1$ which is to be adapted for an explicit solution. 
This is also in line with Ref.~\cite{Chakraborty:2019uxk} which simply sets $c=1$.
Since the up- and charm-quark masses are much lighter than the top-quark mass, we neglect them in the following analytical investigation.
All quark masses have been taking into account in the numerical investigation of Sec.~\ref{sec:numericalInvest}. 
Since the down-type quarks are much less hierarchical than the up-type quarks, we choose to ignore CKM mixing effects in $m_D$ and assume that the CKM matrix mainly stems from the down-quark mass matrix.
Anyway, it would be a small effect on our calculations that we can safely ignore.
In the RH neutrino mass basis, the masses in the type-I sector are fixed by up-type quark masses, the ratio between the LH and RH energy scales and low-energy neutrino mixing parameters: 
\begin{align}
    m_{I} \simeq - r\, M_{D}^{T}\, D_{\nu}^{-1}M_{D}\,, \mbox{ where }
    M_{D}=U^{T}\, m_{D}\, .
\end{align}
Inserting Eq.~\eqref{eq:DiracNuMass} implies that $(m_{I})_{\alpha \beta} 
\simeq (m_{I})_{\tau \tau} \simeq - r\, c^{2} \, m_t^2 \, (U_{\tau i})^2 /m_i$.


\subsection{Type-II dominance}


\begin{figure}[t]
    \centering
    \includegraphics[width=\textwidth]{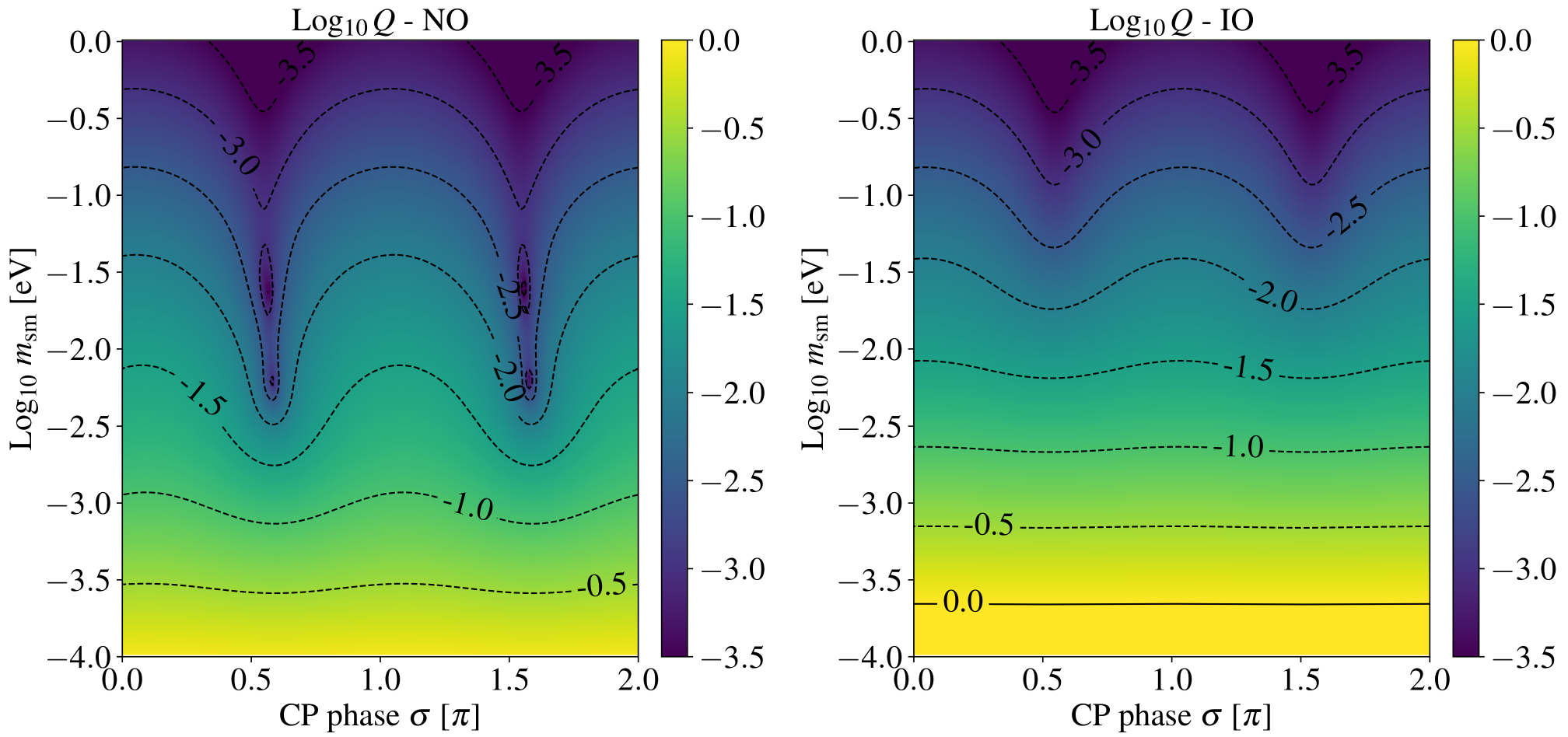}
    \caption{Type-II dominance parameter $Q$ defined in Eq.~\eqref{eq:Qparameter} in dependence of the Majorana phase $\sigma$ and the lightest neutrino mass $ m_{\rm sm}$ for normal (left) and inverted mass ordering (right).
    In order to compare both mass orderings, the two remaining CP phases are set to fixed values: $\delta = 3\pi/2$ and $\tau=0$.
    The RH mass scale is fixed at the GUT scale, $v_{R} =  3\cdot10^{16}$\ GeV.}
    \label{fig:Qparameter}
\end{figure}

\noindent Before we turn to our leptogenesis analysis, we shall demonstrate the consequences of a type-II dominance and, further, formulate the conditions that guarantee the validity of our conclusions.
In order to know when the type-II contribution dominates the light neutrino mass, the
individual seesaw parts have to be estimated.
With our assumptions, the type-I neutrino mass scale is given by
\begin{equation}
\label{eq:typeImassscale}
\overline{m}_{I}\equiv \sqrt{{\rm Tr}\left( {m_{I}}^{\dagger} m_{I} \right)}=r\, c^{2}\, m_{t}^{2} 
\sqrt{ \sum_{i,j} \frac{ U^{2}_{\tau i} U^{2*}_{\tau j} }{m_{i}m_{j}}}
\equiv \frac{r\, c^{2}\, m_{t}^{2}}{\widetilde{m}} \,,
\end{equation}
whereas the type-II neutrino mass scale is 
\begin{align}
\label{eq:typeIImassscale}
\overline{m}_{II} \equiv \sqrt{{\rm Tr} \left({m_{II}}^{\dagger} m_{II}\right)} = 2\,\lambda_{L}\,v_{L} \,,
\end{align}
where $\lambda_L$ was defined in Eq.~\eqref{eq:tripletDecayRates_lepton}.
In our framework, Eq.~\eqref{eq:typeIImassscale} sets the absolute light neutrino mass scale $\overline{m}=\sqrt{m^{2}_{1}+m^{2}_{2}+m^{2}_{3}}$.
With the above expressions, the type-II dominance condition reads 
\begin{align}\label{eq:Qparameter}
Q =\frac{\overline{m}_{I}}{\overline{m}_{II}}\simeq\frac{r\, c^{2}\, m_{t}^{2}}{\overline{m} 
\widetilde{m}}\ll 1\, .
\end{align}
Using the parameter $Q$, the type-II dominance region in the accessible parameter space can be identified.
Namely, we can rewrite the condition $Q \ll 1$ in several insightful ways for
\begin{itemize}
    \item the ratio of $SU(2)_L$ triplet VEV and $m_N$ mass scale:
\end{itemize}
\begin{align}
r\ll \frac{\overline{m}\, \widetilde{m}}{c^{2}\, m_{t}^{2}}\sim 10^{-26} 
\left(\frac{\widetilde{m}}{0.01\ \text{eV}}\right)
\left(\frac{\overline{m}}{0.05\ \text{eV}}\right);
\end{align}
\begin{itemize}
    \item the $SU(2)_L$ triplet VEV $v_{L}$: 
\end{itemize}
\small
\begin{align}\label{eq:LHVev}
v_{L}\ll \frac{\overline{m}\, \widetilde{m}\, v_{R}}{c^{2}\,m_{t}^{2}}\sim 0.1\ \text{eV} 
\left(\frac{\widetilde{m}}{0.01\ \text{eV}}\right)
\left(\frac{\overline{m}}{0.05\ \text{eV}}\right)
\bigg(\frac{v_{R}}{10^{16}\ \text{GeV}}\bigg);
\end{align}
\normalsize
\begin{itemize}
    \item  the trilinear scalar coupling:
\end{itemize}
\small
\begin{align}
\begin{aligned}
\mu &\ll \frac{\overline{m}\, \widetilde{m}\, M^{2}_{\Delta} v_{R}}{c^{2}\, m_{t}^{4}}\sim 
10^{10}\ \text{GeV} \left(\frac{\widetilde{m}}{0.01\ \text{eV}}\right)
\left(\frac{\overline{m}}{0.05\ \text{eV}}\right)\left(\frac{M_{\Delta}}{10^{12}\ 
\text{GeV}}\right)^{2}\bigg(\frac{v_{R}}{10^{16}\ \text{GeV}}\bigg).
\end{aligned}
\end{align}
\normalsize
The philosophy behind these bounds is that if, e.g., the trilinear coupling $\mu$ and the triplet VEV are kept small, the ratio of $v_L$ and the RH neutrino mass scale is small.
Note that $\tilde m$ defined in Eq.~\eqref{eq:typeImassscale} contains flavor dependence, so that the conditions for type-II dominance do not only depend on mass scales.


Another constraint that needs to be taken into account is that the ratio of $v_L$ and the neutrino mass scale should not be too large in order to keep the triplet-lepton coupling matrix $f$ perturbative, 
\begin{align}\label{eq:perturbConstraint}
\lambda_{L}=\frac{\overline{m}_{II}}{2\,v_{L}}\simeq\frac{\overline{m}}{2\,v_{L}}\lesssim
\mathcal{O}(1)\,. 
\end{align}
From now on, we set $v_{L} = \overline{m}/2 = \sqrt{m_1^2 + m_2^2 + m_3^2}/2$, which is close to the smallest possible value that is still consistent with perturbative Yukawa couplings.
This assumption will lead to large Yukawa couplings and triplet decays predominantly into leptons and later on we will see that this eliminates flavor effects in the leptogenesis scenario considered.
At the same time, it fixes the trilinear scalar coupling $\mu$ in terms of the triplet mass $M_\Delta$, eliminates a free parameter from our analysis and allows to rewrite the parameter $\lambda_H$ governing the decay of the triplets into the Higgs, cf.\ Eq.~\eqref{eq:tripletDecayRates_higgs}: 
\begin{align}\label{eq:pertubMu}
\mu = \frac{\overline{m} M^{2}_{\Delta}}{v^{2}_{\rm ew}} \quad \Rightarrow \quad \lambda_{H}=
\frac{\mu}{M_{\Delta}}=\frac{\overline{m} M_{\Delta}}{v^{2}_{\rm ew}}\,.
\end{align}
Together with the bound on $\mu$, this results in a lower bound on $v_{R}$ in dependence of $\widetilde{m}$: 
\begin{align}\label{eq:RHVEVconstr}
v_{R}\gg \frac{v^{2}_{\rm ew}}{4\widetilde{m}}\sim 1.5\cdot 10^{15}\ \text{GeV} 
\left( \frac{0.01\ \text{eV}}{\widetilde{m}}\right). 
\end{align}
As expected this lies around the GUT scale and demands $\widetilde{m}$ not to be too 
small, which is in line with our observations above. 


The behavior of the parameter $Q$ defined in Eq.~\eqref{eq:Qparameter} under assumption of maximally allowed couplings $f$ is illustrated in Fig.~\ref{fig:Qparameter} for certain parameter values and both mass orderings.
For simplicity, the Dirac and Majorana phases $\delta$ and $\tau$ are set to $3\pi/2$ and zero, while we generically fix the RH VEV at the GUT scale, $v_{R}=3\cdot 10^{16}$\ GeV. 
Then, the ratio of LH and RH scales is given by $r\sim 10^{-29}\, (\overline{m}/0.05\ \text{eV})$, assuming maximal perturbative Yukawa couplings in addition. 

\noindent A general feature, regardless of mass ordering, is that type-II dominance induces lower limits on the lightest neutrino mass, $m_{1}$ for normal mass ordering (NO) and $m_{3}$ for inverted mass ordering (IO), respectively. 
For normal mass ordering, type-II dominance is guaranteed for $m_{1}\gtrsim 10^{-3.5}$\ eV, whereas inverted mass ordering favors slightly higher values, $m_{3}\gtrsim 10^{-3.0}$\ eV, when the RH VEV is fixed at the GUT scale,  $v_{R} =  3\cdot 10^{16}$\ GeV.
Note that we have assumed CKM mixing effects to be negligible and, further, $c\sim1$, which might be different in an explicit model building realization.
However, this simply corresponds to a rescaling of our $Q$ parameter which is illustrated in Fig.~\ref{fig:Qparameter}.
Large regions of our framework's parameter space exhibit deviations of a factor of $\sim10$ from the generic left-right-symmetric relation between the VEVs, i.e.\ $v_{L} v_R\sim v_{\mathrm{ew}}^{2}$, which we find acceptable.
We do not specify the origin of this factor of $10$ and leave its explanation in concrete UV completions for future works.
Further, the plot shows that certain phase configurations can lead to an increased type-II contribution for given values of $m_{\rm sm}$. 
This behavior becomes more complex when the other CP phases are varied in addition. In our later numerical analysis, we will make sure that the type-II dominance condition is always fulfilled.
%

\section{Type-II-dominated leptogenesis}
\label{sec:leptogenesis}

\begin{figure}[t]
    \centering
	\includegraphics[width=\textwidth]{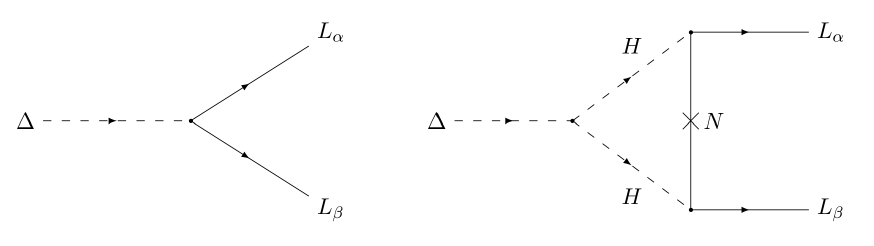}
	\caption{Relevant Feynman diagrams for type-II-dominated leptogenesis. 
	}
	\label{fig:FeynmanDiagram}
\end{figure}

\noindent We will now investigate the capability of the scenario to create the observed BAU via leptogenesis. 
We rely on a typical thermal scenario, where a heavy particle, here the scalar triplet, decays at a temperature around its mass and creates a lepton asymmetry that is later transferred into a baryon asymmetry via non-perturbative SM sphaleron processes.
Pure type-II leptogenesis with only one scalar triplet is not possible because of missing self-energy or vertex diagrams that are needed to interfere with the tree-level decay diagram.
However, our mixed type-(I$+$II) framework allows for a valid scenario, as the heavy RH neutrinos induce a vertex correction that interferes with the tree-level amplitude, cf.\ Fig.~\ref{fig:FeynmanDiagram}, leading to a nonzero CP asymmetry. 
In principle, also these SM singlets decay and could contribute to the creation of a lepton asymmetry, but type-II dominance forces them to be much heavier than the scalar triplet, $M_{i}\gg M_{\Delta},\ i=1,2,3$.
Hence, they decay much earlier and the created asymmetries get washed out in the primordial thermal bath before the triplets start decaying.
The same reasoning applies to potential CP-violating RH triplet decays $\Delta_{R}$, whose masses are assumed to lie around the GUT scale.
In the following investigation, we will assume the scalar triplet to be solely responsible for the creation of the observed BAU, which justifies integrating out the RH singlets. We therefore have a situation in analogy to the one discussed in  Ref.~\cite{Lavignac:2015gpa}.
In our analysis, we will adopt the density matrix formalism developed in Ref.~\cite{Lavignac:2015gpa} in combination with the conditions for flavor regimes and associated spectator corrections of Ref.~\cite{Sierra:2014tqa}.


\subsection{Type-II-dominated CP asymmetry} \label{sec:TypeIIDomCP}


\noindent Before we are going to list all ingredients that are necessary for a full numerical investigation, we want to focus on the implications of type-II dominance for the CP asymmetry parameter~$\epsilon$.
The CP asymmetry of a certain reaction channel is defined as the difference between the channel's rate and its CP-conjugated one, normalized to the total decay width.
For scalar triplet decays these quantities are defined as\footnote{We follow the convention of  Ref.~\cite{Lavignac:2015gpa}, which includes in particular the factor $2$ in $\epsilon_H$.}
\begin{equation}\label{eq:CPEpsilons}
\begin{aligned}
\epsilon_{\alpha \beta}
&=\frac{\Gamma(\overline{\Delta}\rightarrow L_{\alpha} L_{\beta}) - \Gamma(\Delta
\rightarrow \overline{L}_{\alpha} \overline{L}_{\beta}) }{\Gamma_{\Delta}
+\Gamma_{\overline{\Delta}}}(1+\delta_{\alpha\beta})\, ,\\
\epsilon_{H}
&= 2\, \frac{\Gamma(\Delta\rightarrow H H) - \Gamma(\overline{\Delta}\rightarrow 
\overline{H} \overline{H}) }{\Gamma_{\Delta}+\Gamma_{\overline{\Delta}}}\, ,
\end{aligned}
\end{equation}
\noindent where $\Gamma_{\Delta}=\Gamma_{\overline{\Delta}}= \Gamma(\Delta\rightarrow \overline{L}\overline{L}) + \Gamma(\Delta\rightarrow H H) $.
The resulting CP asymmetry corresponding to the diagrams in Fig.~\ref{fig:FeynmanDiagram}, cf.\ Ref.~\cite{Hambye:2003ka}, is then given by 
\begin{align}\label{CPasymGeneral}
\epsilon_{\Delta} = -\frac{1}{8\pi} \sum_{i} M_{i}\frac{ \text{Im} \left[ \mu \left( y^{*}\,f\,y^{\dagger}  \right)_{ii} \right]}{M^{2}_{\Delta} \text{Tr}\left[ff^{\dagger}\right] + |\mu|^{2} } \ln \left( 1 + \left(\frac{M_{\Delta}}{M_{i}}\right)^{2} \right)\, ,
\end{align}
where we recall that $\mu$ is the trilinear triplet-Higgs-Higgs coupling, $y$ the Dirac Yukawa and $f$ the triplet Yukawa coupling.
Using the mass hierarchy related to type-II dominance implies $M_{i}\gg M_{\Delta}$,
and applying the expressions of the individual seesaw contributions, see Eqs.~\eqref{eq:typeImass} and (\ref{eq:typeIImass}), as well as Refs.\ \cite{Hambye:2005tk,Chakraborty:2019uxk},  we can further simplify, which leads us to the compact formula\footnote{The equality of both CP asymmetry parameters can be understood by comparing the interference parts of both reaction channels, which are similar in couplings up to operations due to different time order. The corresponding bosonic diagrams are obtained by replacing lepton with Higgs lines (and vice versa) in Fig.~\ref{fig:FeynmanDiagram}.}
\begin{align}\label{eq:CPasym}
\begin{aligned}
\epsilon_{\Delta}
&=\sum_{\alpha}\epsilon_{\alpha\alpha}
=\frac{1}{4\pi}\frac{M_{\Delta}}{v^{2}_{\rm ew}\, \overline{m}_{II}}\sqrt{B_{L}B_{H}}
\, \text{Im} \left[ {\rm Tr}\! \left[m_{II}{m_{I}}^{\dagger} \right] \right]=\epsilon_{H} .
\end{aligned}
\end{align}
The advantage of type-II dominance is that, in this expression, $m_{II}$ equals the neutrino mass matrix $m_\nu$. 
Furthermore, using a discrete left-right symmetry and hierarchical Dirac Yukawa couplings means that $m_I = - m_D^T \, m_N^{-1} \, m_D$ simplifies because of $m_N \propto m_{II}$ and $m_D \simeq (0,0,m_t)$, cf.\ Eq.~\eqref{eq:typeImass}.
Evaluating the trace as well as the geometric mean of the branching ratios for $v_{L}=\overline{m}/2$, cf.\ Eq.~\eqref{eq:perturbConstraint}, allows us to
re-express $\epsilon$ in a form that factorizes nicely into low- and high-energy quantities:  
\begin{subequations}\label{eq:CPseparated}
\begin{gather}
\hfill \epsilon_{\Delta}= -A\cdot B\ , \quad \text{with } \hfill\\
A=\sum_{i,j}\frac{m_{i}}{m_{j}} \, \text{Im}\! \left[ \left( U_{\tau i} U_{\tau j}^{*}\right)^{2}
\right] ,\quad
B=\frac{\lambda_{H}}{1+\lambda^{2}_{H}}\frac{M_{\Delta}}{4\pi v_{R}}\, ,
\end{gather}
\end{subequations}
where we used Eq.~\eqref{eq:perturbConstraint}. 
We see that $A$ only contains low-energy flavor and CP parameters, whereas $B$ only contains high-energy quantities. 

Actually the CP asymmetry in Eq.~\eqref{eq:CPasym} is the unflavored one, $\epsilon_\Delta = {\rm Tr}\, \epsilon_{\alpha\beta}$.
The flavored asymmetries read in general~\cite{Lavignac:2015gpa}
\begin{align}\label{eq:eps_flav}
	\epsilon_{\alpha\beta}=-\frac{1}{8\pi i}\frac{M_{\Delta}}{v^{2}_{\rm ew}}
    \frac{\sqrt{B_{L}B_{H}}} {\overline{m}_{II}}
    \left(m_{I} {m_{II}}^{\dagger} - m_{II}{m_{I}}^{\dagger}\right)_{\alpha\beta} .
\end{align}
As it turns out, for our purposes, flavor effects (to be discussed in what follows Sec.~\ref{sec:LeptoFlavor}) play only a minor role. 
Hence the unflavored formalism discussed here allows a straightforward investigation of the available parameter space.

First, note that $A$ and hence the BAU depend on the low-energy CP phases $\delta$, $\sigma $ and $\tau$. 
To obtain a better understanding of the dependence on low-energy observables, we further simplify  $A$. 
We introduce two dimensionless auxiliary quantities $R$ and $\eta$ to take into account various possible mass orderings and hierarchies: $\delta m^{2}=R \, |\Delta m^{2}|$ and $m_{\rm sm}^{2}=\eta \, |\Delta m^{2}|$.
By fixing the remaining mixing angles\footnote{Here, we set $\theta_{23}=\frac{\pi}{4}$ and $\sin^{2} \theta_{12} =\frac{1}{3}$.} 
and keeping terms up to first order in $R$ and $\theta_{13}^{2}$, we obtain compact expressions that allow a qualitative understanding of how accessible low-energy observables influence the creation of a CP asymmetry:
\begin{equation}\label{eq:analyticA}
\begin{aligned}
A&=S_{1}+\frac{T S_{2}}{96\sqrt{\eta(\eta +1)^{3} }} \\
&+ \frac{(\eta(\eta +1))^{-3/2}}{192}
\begin{cases}
2 (\eta (T - R ) - 2 R) \sin2(\sigma +\tau)
    +\eta(T+R)\sin 2\tau \, , & \quad \text{NO}\, ,\\
2\eta (T+R)\sin 2(\sigma +\tau )
    +\eta (T-R)\sin 2\tau\, , & \quad \text{IO}\, ,\\
\end{cases}
\end{aligned}
\end{equation}
\noindent where the full dependence on $\delta$ is incorporated in the parameters $S_{1,2}=S_{1,2}(\delta, \sigma,\tau, \theta_{13}^{\,}, \theta_{13}^{\,2})$, while $T=T(\eta)$ only depends on the introduced auxiliary parameters $\eta$. 
The full expressions for both terms are given in Eqs.~\eqref{eq:appS} in \ref{app:AppB}.\footnote{The analytical calculations throughout this work are done with the computer algebra system Mathematica~\cite{Mathematica}.}
Using these expressions together with Eqs.~\eqref{eq:analyticA}, one can deduce that the leading dependence on the Dirac phase goes with $\sin \delta$, and with  
$\sin 2 \alpha$ and higher periodicities for the Majorana phases $\alpha = \sigma, \tau$.
Finally, we can approximate $R\simeq\frac{1}{30}$ and $\theta_{13}\simeq\frac{1}{7}$, which leaves us with the following approximation of the parameter $A$ in case of normal neutrino mass ordering and a lightest neutrino mass around $10^{-2}$ eV (in this case $R$ and $\eta$ are of similar magnitude)
\begin{subequations}
    \begin{equation}
    \small{
    \begin{aligned}
     A(m_{1}&=10^{-2}\ \text{eV}| \sigma,\tau=0) = \frac{\theta_{13}^2}{96} \bigg[16-3 \sqrt{2}+\frac{\sqrt{2}+4}{\sqrt{R}}\bigg] \sin 2 \delta \\
     &+\frac{\theta_{13}}{288} \bigg[\left(33-21 \sqrt{2}\right)\sqrt{R} +4 \left(2 \sqrt{2}-3\right) - \frac{12}{\sqrt{R}} \left(\sqrt{2}-1\right)\bigg] \sin\delta \, ,
    \end{aligned}
    }
    \end{equation}
    \begin{equation}
    \small{
    \begin{aligned}
     A(m_{1}&=10^{-2}\ \text{eV}|\delta=\pi)  = \frac{\widetilde{A}_{1}}{576} + \frac{\theta_{13}^{2}}{96} \bigg[-3 \sqrt{2} \sin 2\sigma + 8 \sin 4\sigma -12 \sin 4\tau \\&- \frac{\sqrt{2}}{\sqrt{R}} \sin 2(\sigma +\tau )  + \frac{2}{\sqrt{R}} \sin 2\tau \bigg] + \frac{\theta_{13}}{288} \bigg[12 \sin 2\sigma - 16 \sqrt{2} \sin 4\sigma \\& +\frac{3 \sqrt{2} (7 R+4)}{\sqrt{R}} \sin 2\tau  -\frac{3 (11 R+4)}{\sqrt{R}} \sin 2(\sigma + \tau) \bigg]\, ,
    \end{aligned}    
    }
    \end{equation}
    \begin{equation}
    \small{
    \begin{aligned}
    A(m_{1}&=10^{-2}\ \text{eV}|\delta=3\frac{\pi}{2})= \frac{\widetilde{A}_{1}}{576} + \frac{\theta_{13}^{2}}{96} \bigg[-\frac{3}{\sqrt{R}} \left(\sqrt{2} \sin 2 (\sigma +\tau ) +2 \sin 2\tau \right)\\& +3 \sqrt{2} \sin 2\sigma -8 \sin 4\sigma
    -12 \sin 4\tau \bigg] +\frac{\theta_{13}}{288} \bigg[3 \sqrt{R} \left(7 \sqrt{2} \cos 2\tau -11 \cos 2(\sigma +\tau ) \right)\\& +  12 \cos 2\sigma +8 \sqrt{2} (1-2 \cos 4\sigma)
   +\frac{12}{\sqrt{R}} \left(\sqrt{2} \cos 2\tau -\cos 2(\sigma +\tau) \right)\bigg]\, ,
    \end{aligned}
    }
    \end{equation}
\end{subequations}
with $\widetilde{A}_{1}=\bigg[\frac{3 \sqrt{2} (11
   R+4)}{\sqrt{R}} \sin 2(\sigma +\tau )+ \frac{3(7 R+4)}{\sqrt{R}} \sin 2\tau  +12 \sqrt{2} \sin 2\sigma +16 \sin 4\sigma +36 \sin 4\tau \bigg]$.

\normalsize
\noindent An approximation for smaller $m_{\rm sm}$ values and corresponding expressions for the inverted neutrino mass ordering are listed in \ref{app:AppB}.
More complicated expressions in matrix form are obtained when flavor effects are present. 
Here, we stay brief as we will demonstrate in the next section that flavor effects are insignificant.


\subsection{Importance of flavor} \label{sec:LeptoFlavor}


\noindent Since we later want to scan over several temperature regimes in the early Universe, we have to consider the individual processes that are in equilibrium, as they may influence the generation of the $B-L$ asymmetry and hence the final BAU.
For instance, lepton Yukawa interactions that reach equilibrium at characteristic temperatures, $T_{\tau}\sim 10^{12}$ GeV and $T_{\mu}\sim 10^{9}$ GeV respectively~\cite{Abada:2006fw,Nardi:2006fx}, will have an impact on the structure of the Boltzmann equations.
Correspondingly, the form of the flavored asymmetries $\epsilon_{\alpha\beta}$ is affected, too.\footnote{The individual temperature regimes can be deduced from the temperature-dependent rate of the muon and tau Yukawa interactions, cf.\ Eq.~\eqref{eq:leptonYukawa}. The electron Yukawa, reaching equilibrium at $T_{e}\sim 10^{5}$ GeV, does not affect the system of Boltzmann equations as the coherence among the three lepton families has already been broken.} 
Leptons produced in decay processes propagate as coherent superpositions of lepton flavors as long as the corresponding lepton Yukawa interaction rates are slower than the expansion of the Universe.
In this sense the unflavored CP asymmetry $\epsilon_\Delta $ in Eq.~\eqref{eq:CPasym} is only valid at high temperatures when all lepton Yukawa interactions are out of equilibrium.
The decoherence induced by these interactions reaching equilibrium, i.e.\ when $\Gamma_{\tau,\mu}(T)\simeq H(T)$, effectively splits the lepton flavor superposition and separates the evolution of the corresponding lepton number asymmetries, which are then governed by individual Boltzmann equations.
Within the density matrix formalism, which we will use in this work, these decoherence effects drive the corresponding off-diagonal elements in the $B-L$ asymmetry matrix to zero, cf.\ Eq.~\eqref{eq:YukawaCorr}, such that the $(3\times3)$-matrix of  $B-L$ asymmetries $\Delta_{\alpha \beta}$ at highest temperatures is gradually shrunk, via a ($(2\times2)$ + 1) description, to three individual quantities accounting for the asymmetry produced in the three individual lepton flavors.\footnote{It is to be noted that this is just an approximate procedure as the transition regions require a quantum-mechanical treatment that also takes into account effects of partial decoherence, which is beyond the scope of this work.}

While the previous statements are generally valid for both type-I and type-II leptogenesis scenarios, the situation within a type-II framework is more complicated since the scalar $SU(2)_{L}$ triplet exhibits two decay channels and, in addition, undergoes gauge scatterings.
If the triplet's leptonic inverse decay (ID) rate, i.e.\ the rate of two leptons fusing into a scalar triplet, is much faster than any lepton Yukawa  interaction rate, $\Gamma_{\rm ID}\gg \Gamma_{f}$, the triplet inversely decays before Yukawa interactions can take place, which effectively suppresses flavor effects until both interaction rates are of comparable size, $\Gamma_{\rm ID}\sim \Gamma_{f}$.
The appearance of decoherence is then delayed until inverse decays into leptons are slower than the lepton Yukawa interactions~\cite{Sierra:2014tqa}:
\begin{align}\label{eq:newFlavorCond}
    \Gamma_{f}\geq B_{L}\,\Gamma_{\Delta}\, \frac{Y^{\rm eq}_{\Delta}}{Y^{\rm eq}_{L}}, \quad \text{for } f=\tau, \mu\, . 
\end{align}
Here $Y^{\rm eq}_{\Delta}$ and $Y^{\rm eq}_{L}$ denote the triplet and lepton equilibrium abundances $Y^{\rm eq}_{x}=n^{\rm eq}_{x}/s$, respectively. 
This condition can be translated into triplet mass bounds indicating what flavor treatment is to be applied~\cite{Sierra:2014tqa}:
\begin{equation}\label{eq:flavorCondStrong}
\begin{aligned}
    M_{\Delta} &\lesssim 4\ \left(\frac{10^{-3}\ \text{eV}}{\widetilde{m}^{\rm eff}_{\Delta}}\right) \cdot 10^{11}\ 
    \text{GeV} , \quad &\text{2-flavor-regime}\, ,\\
    M_{\Delta} &\lesssim 1\ \left(\frac{10^{-3}\ \text{eV}}{\widetilde{m}^{\rm eff}_{\Delta}}\right) \cdot 10^{9}\ 
    \text{GeV   } , \quad  &\text{3-flavor-regime}\, ,
\end{aligned}
\end{equation}
with the effective triplet mass parameter $\widetilde{m}^{\rm eff}_{\Delta} = \overline{m}_{II} \sqrt{(1-B_{H})/B_{H}} =\left( \lambda_{L}/\lambda_{H}\right) \overline{m}_{II}$.
So in fact, two conditions have to be fulfilled for flavor effects to be relevant: first, the necessary condition that a certain lepton Yukawa interaction reaches equilibrium, $\Gamma_{f} \simeq  H$, and second, the sufficient condition that it is faster than inverse triplet decays given by Eq.~\eqref{eq:newFlavorCond}.
Our previous assumptions of type-II dominance and perturbativity of $\lambda_{L}$ lead to a characteristic expression of the effective triplet mass parameter, $\widetilde{m}^{\rm eff}_{\Delta} \simeq v^{2}_{\rm ew}/M_{\Delta}$, such that the upper mass bounds in Eqs.~\eqref{eq:flavorCondStrong} are shifted towards energies that are irrelevant within a thermal leptogenesis scenario like it is assumed in this paper.
In particular, for a typical triplet mass of $M_{\Delta}\sim10^{10\,-\,13}$~GeV we obtain $\widetilde{m}^{\rm eff}_{\Delta} \simeq 6\cdot 10^{3\,-\,0}$~eV, which indicates that flavor effects will play a minor role in the following investigations.

Finally, we want to comment on the gauge interactions related to the weak charge carried by the scalar $SU(2)_{L}$ triplet.
These fast gauge boson scatterings inhibit sufficient asymmetry production until the associated rate drops below the usual triplet decay rate, $\Gamma_{A} < \Gamma_{D}$.
Thus, effective BAU generation is restricted to times when gauge scatterings proceed sufficiently slower than triplet decays.
Since these bounds from gauge scatterings are generally assumed to be weaker than the ones from Eq.~\eqref{eq:newFlavorCond}, cf.\ Ref.~\cite{Sierra:2014tqa}, we only apply the previous flavor condition, cf.\ Eq.~\eqref{eq:newFlavorCond}, and shift a detailed discussion of these "weak" flavor conditions for the interested reader to \ref{app:AppC}.

In summary, the fact that for our set of parameters the triplet couples more strongly to leptons, $\lambda_H \ll \lambda_L = 1$, leads to a fast inverse-decay rate $\Gamma_{\rm ID}$, which implies that flavor effects are not very significant. 
Nevertheless, in the course of this work we will carefully consider the different flavor regimes that are related to flavor effects and spectator processes, which is an interesting analysis in its own right.
Note that the notion of "unflavored leptogenesis" is not as easy to define as in the type-I seesaw case. We return to this point in Sec.~\ref{sec:1vs3}.


\begin{table}[t]
\caption{
    Model parameters relevant for the evolution of particle abundances and the quantification of the individual washout terms, see Fig.~\ref{fig:AsymEvol} and Tab.~\ref{tab:WashoutComp}.
    We also refer to these values as default when individual parameters are varied in the following discussion.
    For comparison, the resulting neutrino mass parameters are listed as well.
    }
    \medskip\centering
    \resizebox{\textwidth}{!}{
    \begin{tabular}{l|c||l|c}
        parameter & default value & neutrino mass parameter & value (NO/IO) \\ \hline
         Smallest neutrino mass $ m_{\rm sm}$ &  1 meV & 
         $Q$ parameter & 0.09 / 0.22 \\  
         Dirac phase $\delta$ &  $\frac{6}{5}\pi$ (NO) /  $\frac{8}{5}\pi$ (IO) & 
         mass scale $\overline{m}$ & 50.0 / 70.2 meV\\ 
         Majorana phase $\sigma$ &   $\pi/2$,  & 
         eff.\ Majorana mass $m_{ee}$ & 2.48 / 18.8 meV\\ 
         Majorana phase $\tau$ & $\pi/2$, & 
          eff.\ el.\ neutrino mass $m_{\nu_{e}}$ & 8.88 / 49.0 meV\\
         triplet mass $ M_{\Delta}$ & $1.2$ (NO) / $0.64$ (IO) $[10^{13}$\ GeV] & 
         $\sum_{i} m_{i}$ & 59.9 / 100.3 meV
    \end{tabular}
    }
    \label{tab:fixedParameters}
\end{table}



\subsection{Numerical investigation}\label{sec:numericalInvest}


\noindent We are now going to perform a full numerical analysis\footnote{The numerical investigation is performed within the SciPy framework~\cite{ScientificComputing2007,ScientificComputing2011,Scipy2020,Matplotlib2007,Ipython2007,Numpy2020,Pandas2010,JupyterLab}, while MPI for Python~\cite{MPI4Py2005, MPI4Py2008} is used for cluster computations.} within a refined framework that applies a density matrix approach~\cite{Blanchet:2011xq, Lavignac:2015gpa} in combination with the flavor transition conditions of Ref.~\cite{Sierra:2014tqa}.
Further, both modified spectator corrections that come along with this flavor transition conditions and washout effects up to $(2\rightarrow2)$ scatterings are incorporated.
Contributions of lighter quark masses in the Dirac neutrino mass, cf.\ Eq.~\eqref{eq:DiracNuMass}, are taken into account as well.


\subsubsection{Boltzmann equations}\label{sec:boltz}


\noindent The dynamical quantities of scalar triplet leptogenesis are the abundances of triplets, $\Sigma_{\Delta}\equiv (n_{\Delta}+n_{\bar{\Delta}})/s$, Higgs and lepton doublets, $\Delta_{X} \equiv (n_{X}+n_{\bar{X}})/s$ for $ X= H,\,L$, and the $B-L$ charge asymmetry matrix $\Delta_{\alpha\beta}$. Here $s$ is the entropy density. 
Since the triplet is not its own antiparticle, the difference between triplets and antitriplets is assigned a separate quantity, $\Delta_{\Delta}\equiv(n_{\Delta}-n_{\bar{\Delta}})/s$. 
The evolution of each quantity is governed by a Boltzmann equation, but because of hypercharge conservation only three of them are independent, $2\,\Delta_{\Delta}+\Delta_{H}-\Delta_{L} = 0$.
Following the convention in the literature, e.g.\ Refs.~\cite{Hambye:2003ka, Hambye:2005tk, Hambye:2012fh, Sierra:2014tqa, Lavignac:2015gpa}, we eliminate the Higgs doublet abundance, such that leptogenesis is described by%
\footnote{We follow the conventions of Ref.~\cite{Lavignac:2015gpa} with minor modifications regarding lepton Yukawa interactions reaching equilibrium as the temperature decreases, cf.\ Ref.~\cite{Sierra:2014tqa}. The appropriate spectator corrections are listed in Tab.~\ref{tab:spectatorCorr}.}
\begin{subequations}\label{eq:BEs}
\begin{align}
	sHz\frac{d\Sigma_{\Delta}}{dz}
	&=-\left(\frac{\Sigma_{\Delta}}{\Sigma^{\rm eq}_{\Delta}}
    -1\right)\gamma_{D} -2\left(\left(\frac{\Sigma_{\Delta}}{
    \Sigma^{\rm eq}_{\Delta}}\right)^{2}
    -1\right)\gamma_{A} \, ,\\ 
    sHz\frac{d\Delta_{\alpha\beta}}{dz}
    &=-\left(\frac{\Sigma_{\Delta}}{\Sigma^{\rm eq}_{\Delta}}-1\right)
    \gamma_{D}\, \epsilon_{\alpha\beta}+W^{D}_{\alpha\beta} + W^{lH}_{\alpha\beta}
    + W^{4l}_{\alpha\beta} + W^{l\Delta}_{\alpha\beta}
    -C_{\tau}-C_{\mu} \, ,\\
	sHz\frac{d\Delta_{\Delta}}{dz}
	&=-\frac{1}{2}\left({\rm Tr}\left(W^{D}_{\alpha\beta}\right)
	-W^{D}_{H}\right)\gamma_{D}\, ,
\end{align}
\end{subequations}
with Hubble rate $H$ and time variable $z\equiv M_{\Delta}/T$; 
$\epsilon_{\alpha\beta}$ corresponds to the flavored CP asymmetries in Eq.~\eqref{eq:eps_flav}, $\gamma_{D}$ to the (inverse) decay reaction rate, $\gamma_{A}$ to the reaction rate induced by gauge scatterings of the triplet, and $\Sigma^{\rm eq}_{\Delta}$ to the triplet equilibrium density.  

\noindent
We list the explicit forms of these expressions in \ref{app:AppA}.  
The washout is split into different terms according to their origin and given as matrices in flavor space:
\begin{itemize}
    \item $W^{D}$: inverse lepton decays, Eq.~\eqref{eq:WD};
    \item $W^{H}$: inverse Higgs decays, Eq.~\eqref{eq:WDHiggs};
    \item $W^{lH}$: lepton--Higgs scatterings, Eq.~\eqref{eq:WlH}; 
    \item $W^{4l}$:  2-lepton--2-lepton scatterings, Eq.~\eqref{eq:W4l}; 
    \item $W^{l\Delta}$: lepton--triplet scatterings, Eq.~\eqref{eq:WlDelta}. 
\end{itemize}
While the first three terms already arise within an unflavored description, the latter two only occur in a flavored treatment when the explicit flavor change of triplet interactions is tracked.
The $C$ terms account for flavor effects due to charged-lepton Yukawa interactions~\cite{Blanchet:2011xq}.
They correct the effect of lepton Yukawa interactions reaching equilibrium as already mentioned in Sec.~\ref{sec:LeptoFlavor} and allow a transition between different flavor regimes.\footnote{In doing so, effects of \textit{partial} decoherence before and after the transition are neglected. Instead of implementing a sharp cut, we use a smooth transition with characteristic temperatures according to the flavor regimes set by Eq.~\eqref{eq:newFlavorCond}.}


\subsubsection{Spectator corrections}


\noindent In addition to lepton Yukawa interactions, many other reactions reach equilibrium as the Universe cools down and are in principle able to influence the evolution of the created $B-L$ asymmetry.
Although they do not directly affect the densities relevant for leptogenesis, they influence quantities that the considered washout terms rely on, i.e., the asymmetries in Higgs and lepton doublets, $\Delta_{L}$ and $\Delta_{H}$, respectively.
Such kind of reactions are, e.g., quark Yukawa interactions as well as electroweak and strong sphalerons, which become relevant at different temperatures.
This property in combination with their connection to washout explains why they are referred to as "spectator processes"~\cite{Buchmuller:2001sr, Nardi:2005hs}.
In general, their impact is quantified by chemical equilibrium conditions when a certain reaction becomes relevant, i.e., faster than the Universe's expansion rate, $\Gamma \gtrsim H$.
As temperature drops, more and more interactions enter equilibrium and contribute new or modify existing conditions.
These chemical equilibrium conditions can be linked to particle--antiparticle asymmetries of the corresponding particle species.
For small chemical potentials, $\mu_i \ll T$, we can write
\begin{align}
\Delta_{f} = Y_{f}-\overline{Y}_{f} = \frac{g_{i} T^{2}}{6s}\mu_{i} \,, \qquad 
\Delta_{b} = Y_{b}-\overline{Y}_{b} = \frac{g_{i} T^{2}}{3s}\mu_{i} \,,     
\end{align}
where $g_{i}$ are the degrees of freedom of particle species $i$, and the difference between fermions $f$ and bosons $b$ is responsible for a factor $1/2$.
All relevant asymmetry densities can then be expressed as linear combinations of $B\!-\!L$ charge densities, $\Delta_{\alpha\beta} = 1/3\,\Delta_{B}\delta_{\alpha \beta} - \left(\Delta_{L}\right)_{\alpha\beta}$. 
In this way, the equilibrium conditions can be used to re-express asymmetries in lepton
and Higgs doublets, $\Delta_{L}$ and $\Delta_{H}$ respectively, in terms of quantities that are relevant in a certain temperature~regime. 
We apply these spectator corrections by closely following the convention of Ref.~\cite{Lavignac:2015gpa}, while we take into account that the modified flavor regimes of Eq.~\eqref{eq:newFlavorCond} also affect the formulation of chemical equilibrium conditions, see e.g.\ Ref.~\cite{Sierra:2014tqa}.
The modified spectator corrections applied in our framework can be found in Tab.~\ref{tab:spectatorCorr} of \ref{app:AppC}.
By considering delayed flavor effects also in spectator corrections and embedding everything in a density matrix formalism we are able to perform a state-of-the-art analysis of our leptogenesis framework.

Our whole discussion about delayed flavor transitions due to fast (leptonic) inverse decays relies on the assumption that the corresponding rate is the same for all lepton flavors.
A more precise treatment would be to explicitly distinguish lepton flavor in the triplet's leptonic inverse decays of Eq.~\eqref{eq:newFlavorCond} as well.
The procedure would be similar to the one presented in Sec.~\ref{sec:LeptoFlavor} and might also affect the spectator corrections of Tab.~\ref{tab:spectatorCorr}.
However, as we emphasized the insignificance of flavor under the assumptions mentioned above, this would be beyond the scope of this work. 


\subsubsection{Assumptions and final baryon asymmetry}


\noindent For our numerical investigation, we shall assume thermal initial conditions for the isospin triplet density $\Sigma_{\Delta}$, while the $B-L$ abundance and residual  abundance $\Delta_{\Delta}$ are set to zero.
The zero net asymmetry of the thermal plasma is justified by the fact that the separation between the type-I and type-II mass scales is assumed to be large, such that a potential asymmetry emerging from RH neutrino decays is washed out in the primordial plasma. 
To guarantee this, we fix the RH neutrino mass scale around the GUT scale, $v_{R} = 3\cdot10^{16}$\ GeV, which is also in concordance with our discussion of type-II dominance, i.e., Eq.~\eqref{eq:RHVEVconstr}. 
Recall that our choice $v_{L}\simeq\overline{m}/2$ is motivated by perturbativity of the Yukawa coupling $f_{\alpha \beta}$, cf.\ Eq.~\eqref{eq:perturbConstraint}, such that the space of unknown parameters is only spanned by the triplet mass $M_\Delta$, the smallest neutrino mass $m_{\rm sm}$, the neutrino mass ordering  and the three leptonic CP phases $\delta$, $\sigma$ and $\tau$.
The triplet mass $M_{\Delta}$ is the only quantity related to high-energy scales, while the remaining quantities can in principle be determined, or at least constrained, at low energies in current or future experiments.
The Boltzmann equations in Eqs.~\eqref{eq:BEs} incorporate a treatment of flavor effects in terms of density matrices that is valid at different temperature regimes.
Further, washout effects up to ($2\rightarrow2$) scatterings, cf.\ Eqs.~\eqref{eq:WlH}, \eqref{eq:W4l} and \eqref{eq:WlDelta}, are taken into account with spectator corrections applied for the relevant temperature regimes, cf.\ Tab.~\ref{tab:spectatorCorr}.
The system of differential equations is solved, and the final baryon asymmetry at $z\simeq 100$ is converted into the observable baryon-to-photon ratio~\cite{Khlebnikov:1988sr, Harvey:1990qw},
\begin{align}\label{eq:BAUcalc}
\eta_B = 7.04 \ \Delta_{B}= 7.04  \cdot \frac{12}{37}\,\Delta_{B-L}= 7.04 \cdot \frac{12}{37}\, {\rm Tr}\left(\Delta_{\alpha\beta}\right) .
\end{align}
This value is then compared to the observed value, $\eta_B^{\rm obs}=\left(6.13\pm0.04\right)\cdot 10^{-10}$, according to the 2018 data release by the Planck collaboration~\cite{Aghanim:2018eyx}. 


\section{Results}
\label{sec:results}


\noindent After setting up our framework, we are now going to present our investigation in several steps.
First, our assumed type-II-dominated leptogenesis scenario is illustrated at a certain parameter point to confirm our conclusions about flavor effects.
Then, we will investigate the influence of selected parameters and check the robustness of our results under variations of the experimental data.
Finally, we will use the observed baryon-to-photon ratio as an additional constraint to link the triplet mass $M_{\Delta}$ to low-energy observables.


\subsection{Time evolution for specific parameter sets}


\noindent We begin with an example scenario to understand the underlying dynamics. 
Choosing certain parameter values that reproduce the observed baryon-to-photon ratio, cf.\ Tab.~\ref{tab:fixedParameters}, we explicitly show the difference between flavored and unflavored asymmetry generation and discuss the contribution of the different washout terms given in Eqs.~\eqref{eq:BEs}.


\subsubsection{One-flavor versus three-flavor treatment\label{sec:1vs3}}

\begin{figure}[H]
	\centering
	\includegraphics[width=1.0\textwidth]{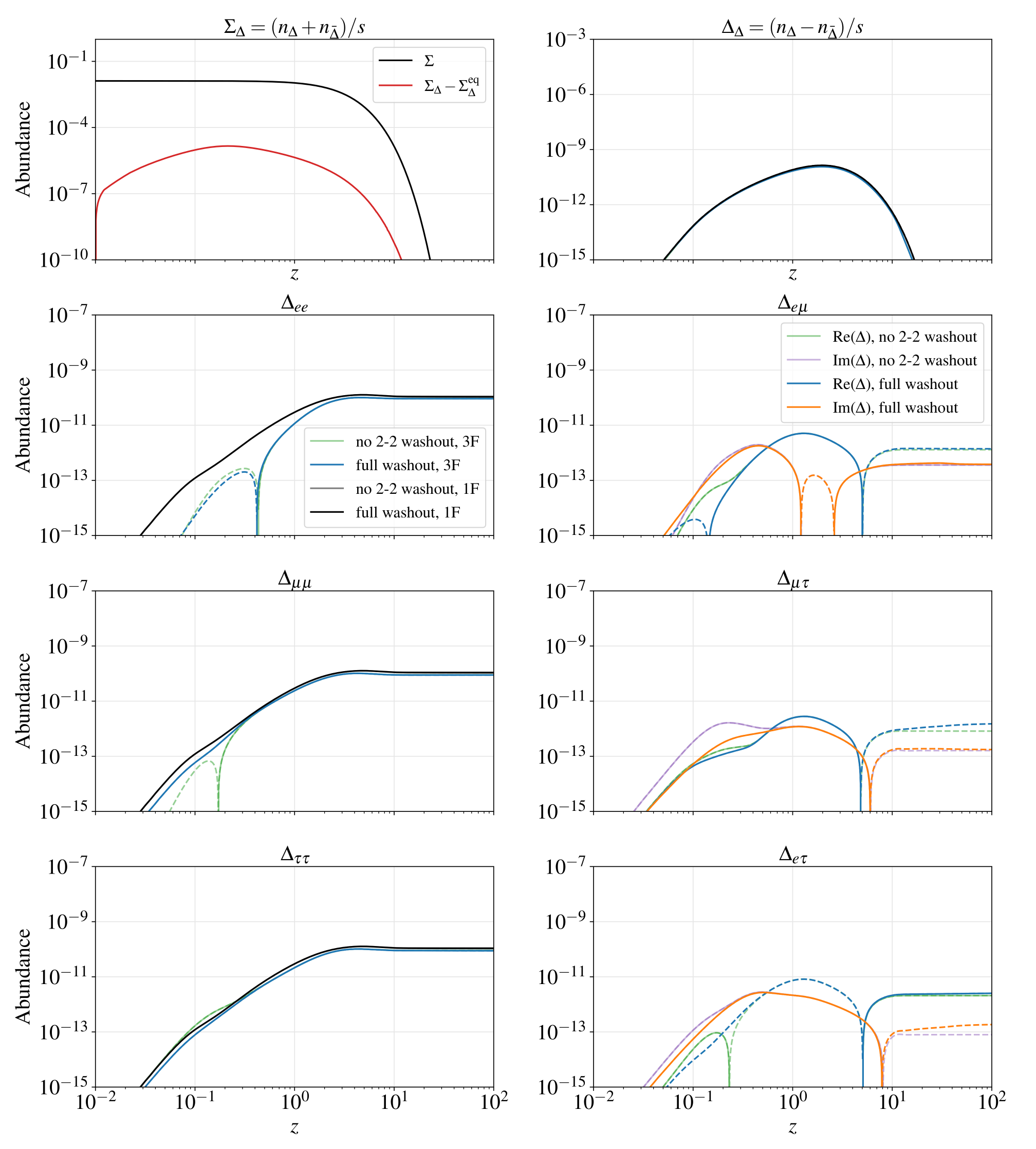}
	\caption{Evolution of scalar triplet abundances, $\Sigma_{\Delta}$ (top left) and $\Delta_{\Delta}$ (top right), as well as $B-L$ asymmetry $\Delta$ as functions of $z = M_{\Delta}/T$. 
	All quantities are shown within a flavored (3F) and an unflavored (1F) framework, respectively.
	For both cases, we distinguish between an evolution taking into account all allowed tree-level washout terms (full washout) and an evolution neglecting washout due to ($2\rightarrow2$) scattering processes (no 2-2 washout).
	In the case of the unflavored $B\!-\!L$ asymmetry, the obtained value is compared to the corresponding diagonal entries of the density matrix $\Delta_{\alpha\alpha}$ for $\alpha = e, \mu, \tau$. 
	Since the off-diagonal entries are allowed to develop an imaginary component, we visualize them in the corresponding plots as well.
	Dashed lines generally indicate the absolute value of the quantities listed in the legends.
	We assume normal neutrino mass ordering with the parameter values listed in Tab.~\ref{tab:fixedParameters}.
	}
	\label{fig:AsymEvol}
\end{figure}

\noindent First, let us discuss the differences between our three-flavor framework and the single-flavor approximation.
The latter is only valid at highest energies, $T > 10^{12}$\ GeV, when all lepton Yukawas are out of equilibrium and the asymmetry generated in lepton doublets can be described by one quantity, $\Delta_{L}$.
Our three-flavor framework, by contrast, incorporates these flavor effects, as well as spectator effects, and is designed to work at all temperatures.
As mentioned in Sec.~\ref{sec:LeptoFlavor}, the naive temperature regimes for tau and muon Yukawa interactions, $T\sim 10^{12}$\ GeV and $T\sim 10^{9}$\ GeV, do not apply since the triplets may undergo fast inverse decays.
Thus, the appearance of flavor effects is generally delayed until inverse triplet decays proceed slower than lepton Yukawa interactions, cf.\ Eq.~\eqref{eq:newFlavorCond}. 
The applied density matrix formalism keeps track of all flavor correlations, and the introduced correction terms, cf.\ Eqs.~\eqref{eq:YukawaCorr}, usually eliminate corresponding off-diagonal entries when a lepton Yukawa interaction reaches equilibrium.
A first look at Fig.~\ref{fig:AsymEvol} already confirms that flavor effects are not contributing since  off-diagonal elements are not driven to zero. 
This is in line with our findings of Sec.~\ref{sec:LeptoFlavor} that flavor effects are of minor importance.
Another difference between the flavored and unflavored frameworks is related to washout.
As the one-flavor framework does not differentiate among flavors, processes that change lepton flavor and eventually the corresponding asymmetry are irrelevant.
The opposite applies to the three-flavor framework, where two additional contributions, four-lepton and lepton-triplet scatterings, occur. 
These terms are able to redistribute the asymmetries stored in the different flavors and potentially protect them from further washout. 
Since this opportunity heavily relies on an interplay between certain flavors, it is rather insignificant for our model.
%

\begin{table}[t]
  \caption{Impact of different $(\Delta L=2)$ washout contributions in the one-flavor and three-flavor frameworks, respectively.
    The values are computed for the parameter set given in Tab.~\ref{tab:fixedParameters}.
    Normal neutrino mass ordering is assumed and the strong flavor condition is
    applied, cf.\ Eq.~\eqref{eq:newFlavorCond}.
    Values in brackets represent final BAU values obtained without any flavor correlation.}
    \medskip\centering
    \begin{tabular}{c|c}
        $2\rightarrow2$ washout  & Baryon-to-photon ratio $\eta_B$ [$10^{-10}$]
        \\ \hline
        none  & 6.224 (6.236)\\
        4-lepton and lepton-triplet & 6.224 (6.236)\\ 
        only lepton-Higgs & 6.137 (6.148)\\  
        all & 6.137 (6.148)\\ \hline\hline 
        none (1F) & 7.610\\ 
        lepton-Higgs (1F)  & 7.344\\ 
    \end{tabular}
    \label{tab:WashoutComp}
\end{table}


\newpage
\noindent The full evolution of the three quantities of interest, $\Sigma_{\Delta}$, $\Delta_{\alpha\beta}$, and $\Delta_{\Delta}$, for both frameworks is shown in Fig.~\ref{fig:AsymEvol} for normal neutrino mass ordering and the parameter set is summarized in Tab.~\ref{tab:fixedParameters}.
Dashed lines generally represent the absolute value of the quantities of interest. 
Naively, we do not expect any large difference in the evolution of the triplet abundances $\Sigma_{\Delta}$ and $\Delta_{\Delta}$, apart from spectator corrections in the three-flavor framework (which are in general small).
This is confirmed by the upper plots of Fig.~\ref{fig:AsymEvol}.
The main difference occurs in the generation of the $B\!-\!L$ asymmetry $\Delta$, which is a single number in the one-flavor treatment and a $(3\times3)$ density matrix in the three-flavor case.
Especially, the density matrix $\Delta$ allows inference of correlations between lepton flavors as this is encoded in its off-diagonal elements.
Since $\Delta$ is Hermitian by construction, only the six independent entries are shown and, in the case of the diagonal entries, compared to the simple one-flavor treatment.
The individual diagonal entries are generally smaller since the final asymmetry is distributed among these quantities and additional washout processes are at work.
As already mentioned above, our choice of Yukawa couplings, $\lambda_{L}\simeq1$, renders flavor decoherence unimportant and thus flavor correlations encoded in the off-diagonal elements sustain.
By weakening this assumption, we expect leptonic triplet decays to be less dominant, such that off-diagonal terms are dynamically driven to zero by the action of the introduced counter-terms.
To estimate the impact of flavor correlation in the creation of a lepton asymmetry, we also listed in Tab.~\ref{tab:WashoutComp} the BAU values in the one-flavor approximation (denoted by 1F) and, in addition, plotted the corresponding evolution in Fig.~\ref{fig:AsymEvol} for illustrative purposes.
Note that in triplet decays the final states are always a pair of leptons $L_\alpha$ and $L_\beta$, and there is no linear combination that can be defined as a single flavor final state. 
This is unlike the case of type-I seesaw, where the decay relevant for leptogenesis is $N_i \to L H$, with $L$ being some combination of lepton flavor states.
Following Ref.~\cite{Lavignac:2015gpa}, we assume that only one flavor exists, and solve the Boltzmann equations that ignore flavor and spectator effects, see also Ref.~\cite{Hambye:2005tk}.
Multiplying with a factor of $3$ to account for the presence of three (equally contributing) flavors then gives the approximate one-flavor result for the baryon asymmetry in Tab.~\ref{tab:WashoutComp}. 
Here the factor of 3 is necessary because triplet decays in the one-flavor approximation (unlike RH neutrino decays in the one-flavor approximation) can only be described in a toy model with one fermion species.
Compared to the exact three-flavor result, the effect is, within our model, less than $15\,\%$.


\subsubsection{Contributions from different washout terms}


\noindent Let us now discuss the contributions of different scattering processes to washout, i.e., contributions from ($2\rightarrow2$) reactions.
Besides the usual inverse decays, there is just one additional contribution in the single-flavor approximation, which originates from lepton-Higgs scatterings.
The mediating particle can be either the $SU(2)_{L}$ triplet or the particle related to the type-I contribution, i.e., in our case, the much heavier RH singlet fermion.
In the three-flavor framework, two additional contributions exist, since triplet interactions can shift a produced asymmetry into different lepton flavors.
These contributions are four-lepton scatterings with the triplet in the intermediate state, $W^{4l}$, and lepton-triplet scatterings with a lepton mediator, $W^{l\Delta}$.
The impact of these processes is displayed in Fig.~\ref{fig:AsymEvol} and quantified in Tab.~\ref{tab:WashoutComp} , while the dynamical evolution of the associated reaction densities is depicted in Fig.~\ref{fig:ReactionDensity} of \ref{app:AppA}.
The parameters have been chosen in such a way that the correct BAU is approximately reproduced when all washout components are at work. 
The negligible impact of washout related to flavor-changing $(2\rightarrow2)$ processes, $W^{4l}$ and $W^{l\Delta}$, respectively, is in line with our previous argumentation, cf.\ Sec.~\ref{sec:LeptoFlavor}, that flavor does not play a significant role in our framework.
Thus, also in the three-flavor treatment, $(2\rightarrow2)$ lepton-Higgs interactions are the major contribution to washout in our leptogenesis scenario.
Again, the situation might change, if one loosens our requirement of maximal perturbative Yukawa couplings, cf.\ Eq.~\eqref{eq:perturbConstraint}.

\begin{figure}[t]
	\centering
	\includegraphics[width=0.47\textwidth]{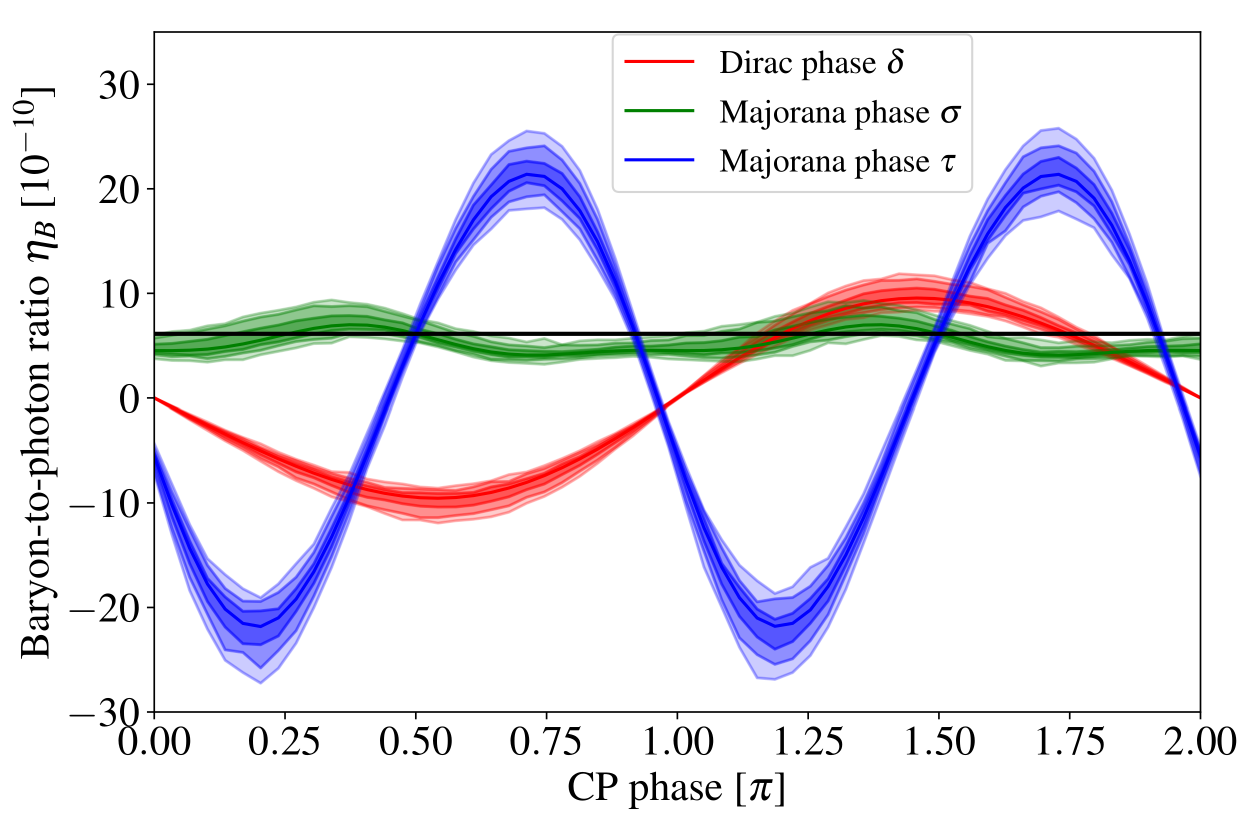}\hfill
	\includegraphics[width=0.48\textwidth]{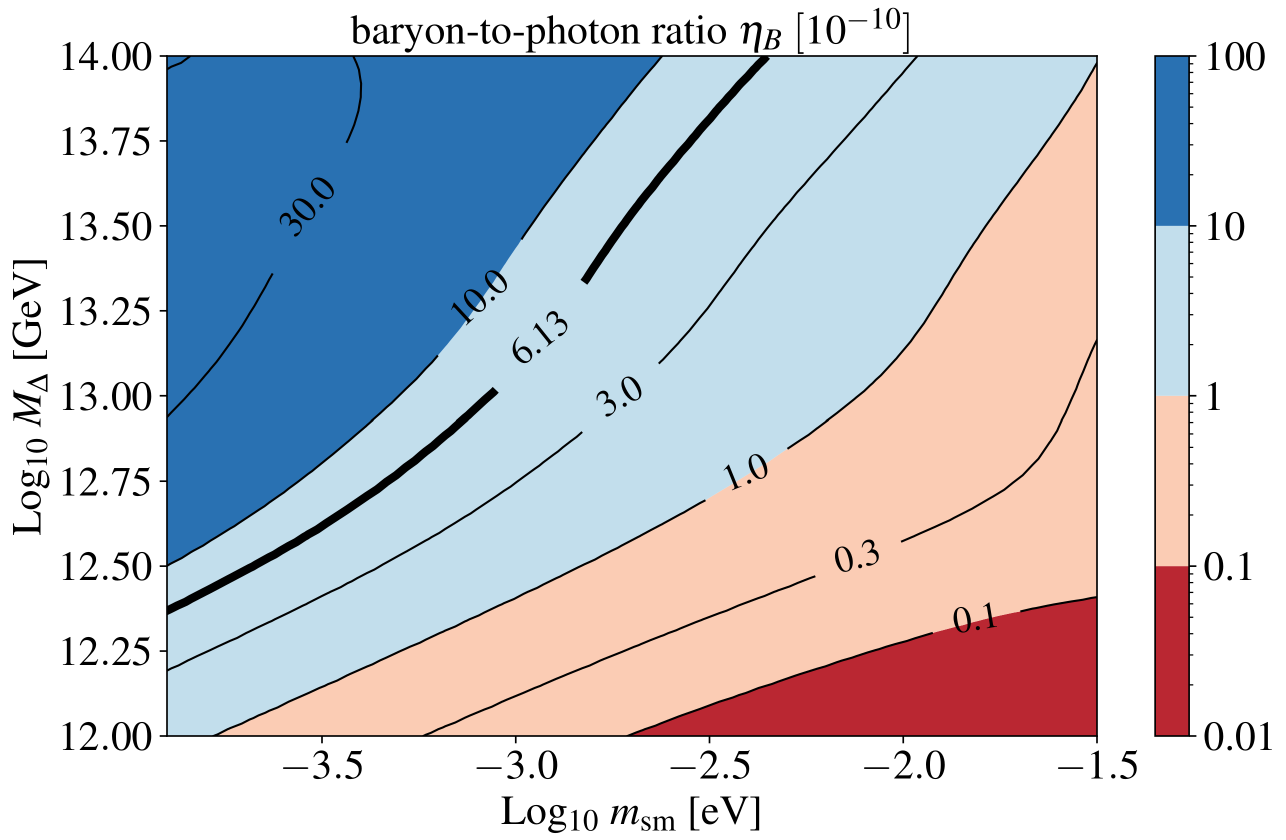}
	\caption{Left: Variation of the baryon-to-photon ratio $\eta_B$ in dependence of the Dirac phase $\delta$ and the Majorana phases $\sigma$, $\tau$ for normal mass ordering.
	The contours reflect the variation of the mixing parameters $\sin^{2}\theta_{12}$, $\sin^{2}\theta_{23}$, $\sin^{2}\theta_{13}$, $\delta m^{2}$, $\Delta m^{2}$ within their experimental uncertainties according to Ref.~\cite{Capozzi:2020qhw}.
    Right: Baryon-to-photon ratio $\eta_B$ in dependence of dimensional parameters, i.e., the lightest neutrino mass $m_{\rm sm}$ and the triplet mass $M_{\Delta}$.
    Normal mass ordering is assumed and the CP phases take the values listed in Tab.~\ref{tab:fixedParameters}.
	\label{fig:CPPhases}
	}
\end{figure}



\subsection{Dimensional parameters and robustness of results}


\noindent Next, let us investigate how the final BAU depends on the low-energy parameters. 
Of particular interest are the three CP phases $\delta$, $\sigma$ and $\tau$. 
Since the latter two realistically show up only in the effective mass for neutrinoless double beta decay~\cite{Rodejohann:2011mu}, 
\begin{equation}
m_{ee} = \bigg| \sum_{i=1,2,3} U_{ei}^2 \, m_i \bigg| \,, 
\end{equation}
we will also plot this quantity instead of $\sigma$ and $\tau$ in the next section.
Before doing so, we want to highlight their individual influence on the resulting BAU.
Hence, we vary them, while fixing all other parameters at the values indicated in Tab.~\ref{tab:fixedParameters}.
Depending on the domain of the varying CP phase, $\eta_B$ changes its sign with a clear difference in periodicity between Dirac and Majorana phases, as discussed above. 
Even more important is that experimental uncertainties on the oscillation parameters can have an effect of up to $25\,\%$.
The right panel of Fig.~\ref{fig:CPPhases} shows $\eta_B$ in dependence of the lightest active neutrino mass $m_{\rm sm}$ and the scalar triplet mass $M_{\Delta}$ in case of normal mass ordering; then $m_{\rm sm}~\widehat{=}~m_{1}$.
A general trend is that, with lighter neutrino mass lighter scalar triplets are needed to obtain the observed BAU.
As indicated by Fig.~\ref{fig:Qparameter}, type-II dominance also depends explicitly on the CP phases.
Thus, the connection between both dimensional quantities change when the CP phases have other values then the ones assumed on Tab.~\ref{tab:fixedParameters}.
Since the effective Majorana mass $m_{\rm ee}$ is strongly altered by the CP phases, the relations between this low-energy observable and the triplet mass $M_{\Delta}$ are more complex and depend on the explicit values, as we will see in the following section.


\subsection{Parameter space scan assuming correct baryon asymmetry}


\noindent Thus far, we have studied the sensitivity of the final BAU on flavor effects and its dependence on the input parameters of our model.
In what follows, we now want to turn this around and use the "condition of successful baryogenesis" 
as an additional constraint to make more quantitative statements about our model's parameter space.
Hence, by fixing two parameters and scanning the remaining ones for their capability to reproduce the observed BAU, we can eliminate unphysical points and establish a connection between some (currently unknown) low-energy observables and physical high-energy model parameters.

\begin{figure}[H]
    \centering
    \includegraphics[width=\textwidth]{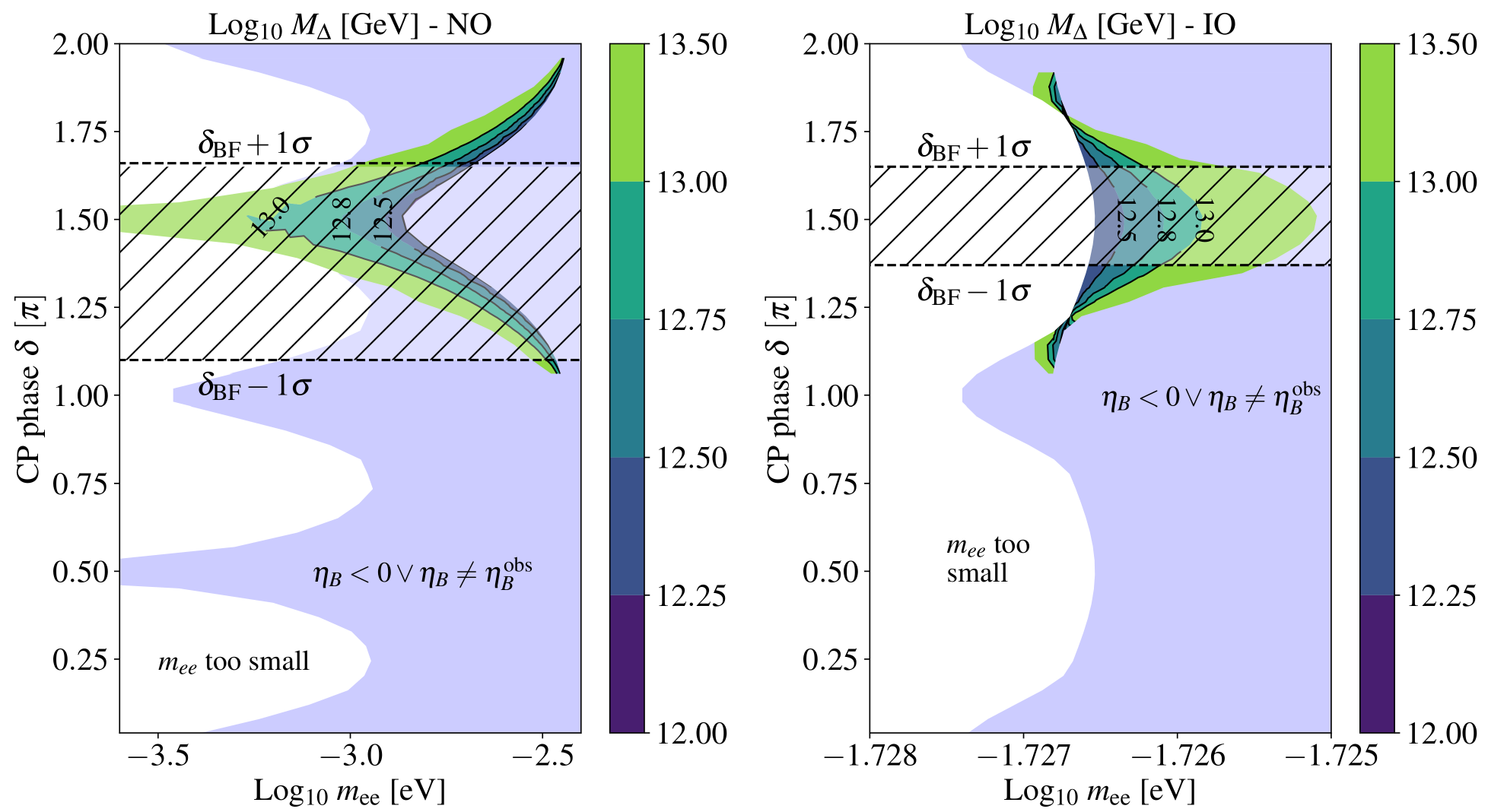}\\
    \vspace{1.5cm}
    \includegraphics[width=\textwidth]{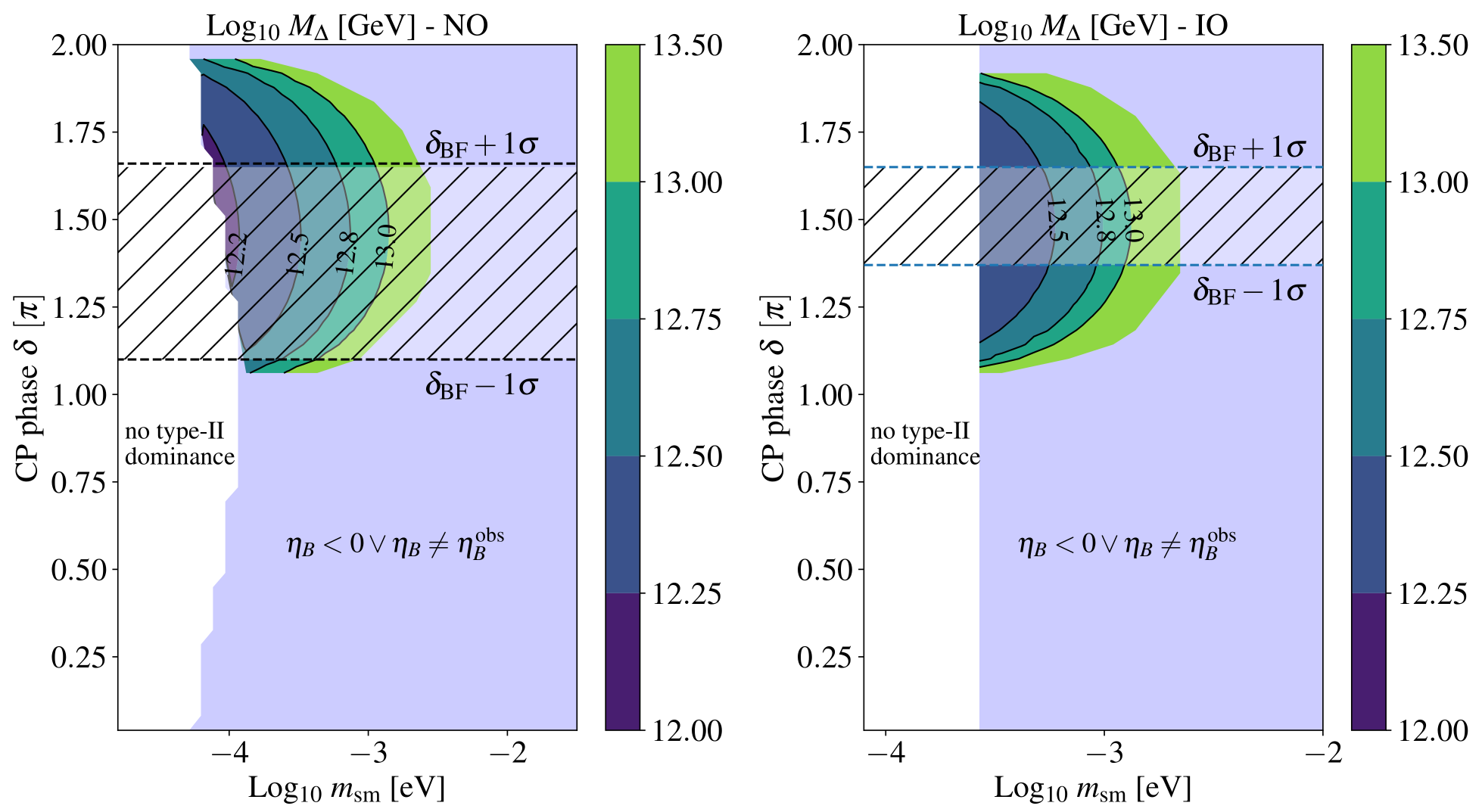}
    \caption{The triplet mass $M_{\Delta}$ as functions of the Dirac CP phase $\delta$ and neutrino mass parameters $m_{ee}$ (top) and $m_{\rm sm}$ (bottom), respectively. 
    We plot all viable points that result in the correct value of the BAU.
    The Majorana phases $\sigma$ and $\tau$ have been set to $\pi/2$, cf.\ Tab.~\ref{tab:fixedParameters}.
    Distinct features are observable for normal (left) and inverted (right) neutrino mass ordering.
    The current best-fit range for the Dirac phase $\delta$ is indicated by the white-shaded band.
    Parameter space regions incapable of producing the observed BAU value are shaded in blue.
    White areas represent parameter space where the effective Majorana mass $m_{\rm ee}$ is too small or our assumption of type-II dominance is not fulfilled.
    }
    \label{fig:BAUconstTripletMass}
\end{figure}

In a first step, we relate the effective neutrino mass $m_{ee}$ and the smallest neutrino mass $m_{\rm sm}$ to the Dirac CP phase $\delta$ and the triplet mass $M_{\Delta}$, cf.\ Fig.~\ref{fig:BAUconstTripletMass}.
The latter one is varied up to $\sim 10^{14}$ GeV ensuring $M_{\Delta}\ll M_{N_{i}}$ and therefore a valid application of Eq.~\eqref{eq:CPasym} is realized.
Parameter values that are not able to reproduce the right baryon-to-photon ratio are shaded in blue, while unphysical points or points that do not fulfill type-II dominance, cf.\ Eq.~\ref{eq:Qparameter}, are left white.
Recent limits on the Dirac CP phase from global fits, cf.\ Ref.~\cite{Capozzi:2020qhw}, are indicated by white hashed regions.

A first insight of this investigation is that Majorana CP phases need to be non-zero.
In case of IO, we can confirm that the Dirac CP phase $\delta$ alone is incapable of producing the right amount of baryon asymmetry for almost all of the remaining parameter space.
For NO, successful leptogenesis is indeed possible but only for $\delta<\pi$, which is in tension with recent global fits that indicate $\delta>\pi$. 
Thus, our model needs non-zero Majorana phases to contribute additional CP violation in case of IO, whereas only a phase shift in the $\delta$-dependence of CP asymmetry is needed for NO. 
To confirm this, we set both Majorana phase is set to a generic value of $\pi/2$, Fig.~\ref{fig:BAUconstTripletMass}.
We immediately recognize that the valid parameter space opens up for IO.
Further, NO populates a parameter space that is now consistent with recent experimental indications for $\delta$. 
The leading behavior (in $\theta_{13}$) of $A$ explains the periodicity in the Dirac CP phase $\delta$, cf.\ Eq.~\eqref{eq:ANOhigh_gen} and Eq.~\eqref{eq:AIOlow_gen} respectively, indicating the need for at least one non-zero Majorana phases to achieve positive BAU values for $\delta>\pi$.
The expected periodicity visible for IO, again induced by the leading behavior of $A$, cf.\ Eq.~\eqref{eq:AIOhigh_gen}, can be confirmed as well.


\newpage
Note the generic behavior of the effective Majorana mass $m_{ee}$ being much higher for IO, cf.\ the upper plots of Fig.~\ref{fig:BAUconstTripletMass}. 
In addition, leptogenesis for IO is only possible within a very narrow mass range setting a minimal value that might be reachable with future experiments.
On the contrary, reproducing the correct BAU implies upper limits on the light neutrino mass parameters $m_{ee}$ for NO. 
Of course, these features are expected to change and possibly more involved when other Majorana phases are assumed. 

\begin{figure}[H]
    \centering
    \includegraphics[width=\textwidth]{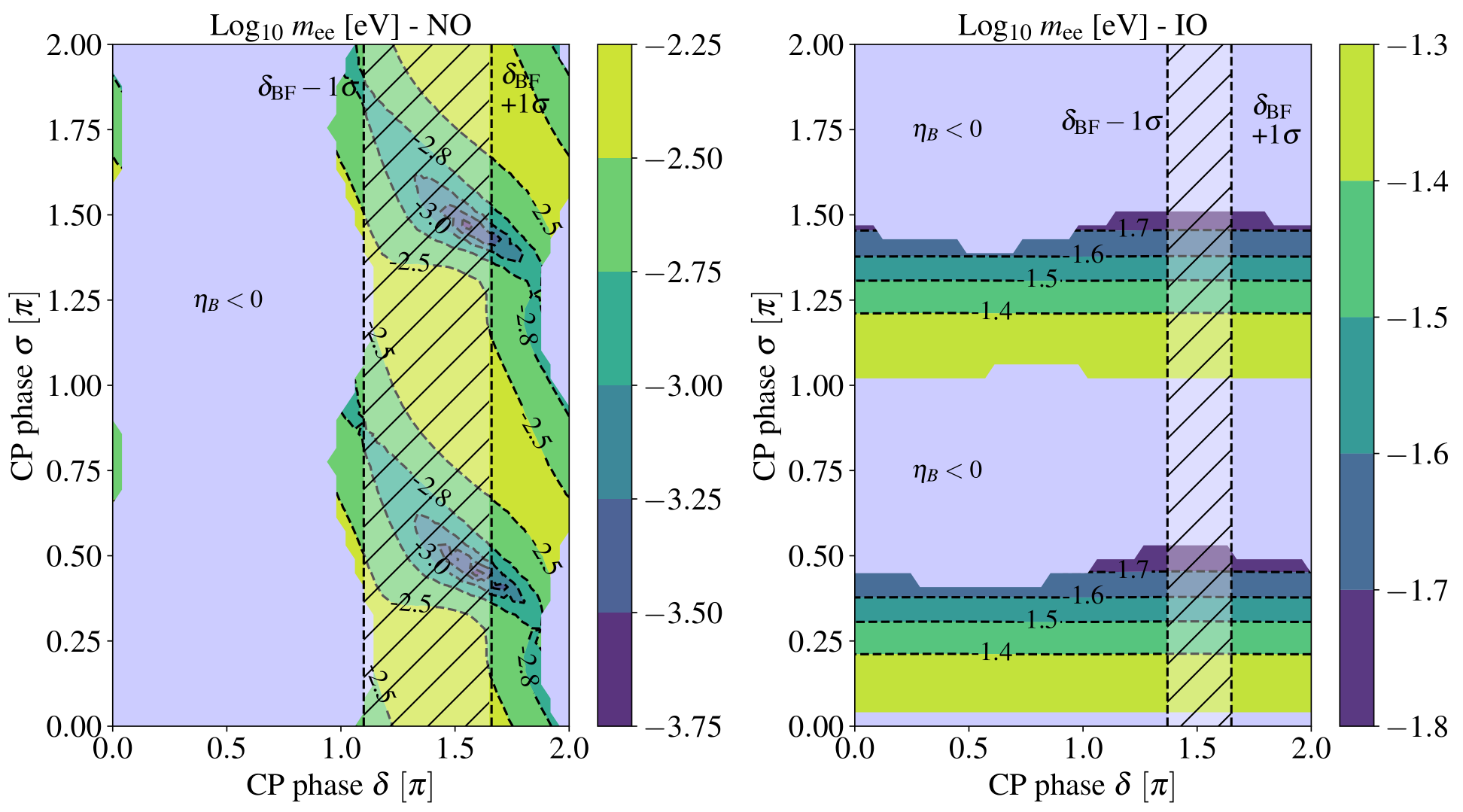}\\
    \vspace{1,5cm}
    \includegraphics[width=\textwidth]{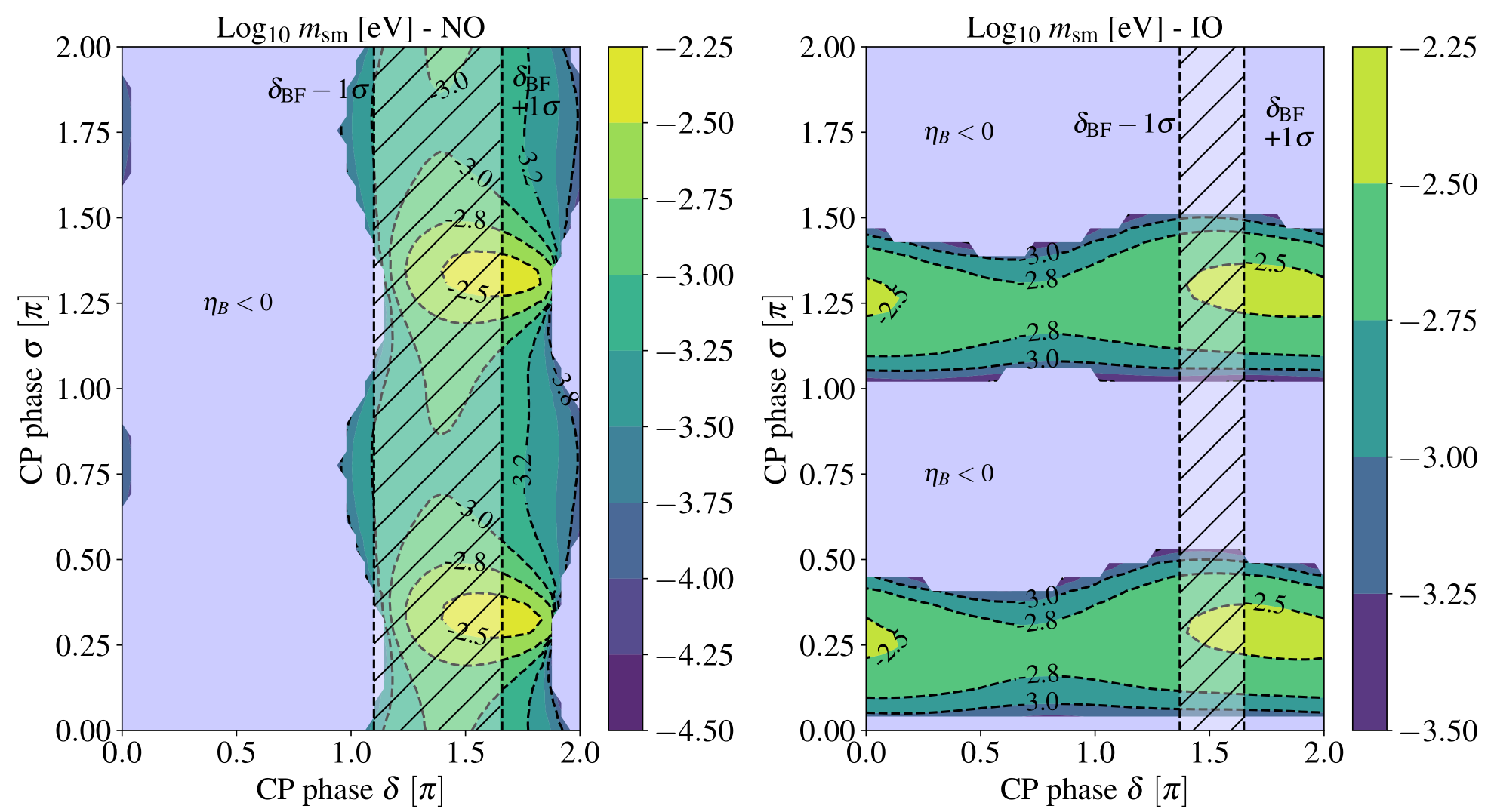}
    \caption{Neutrino mass parameters $m_{ee}$ (top) and $m_{\rm sm}$ (bottom) as functions of the CP-violating phases $\delta$ and $\sigma$. 
    As before, we plot all viable points that result in the correct value of the BAU. 
    The triplet mass is fixed, $M_{\Delta}=1.2\cdot 10^{13}$\ GeV for NO and $M_{\Delta}=0.64\cdot 10^{13}$\ GeV for IO respectively, and the remaining Majorana phase $\tau$ is set to $\pi/2$, cf.\ Tab.~\ref{tab:fixedParameters}.
    Note the distinct features for normal (left) and inverted (right) neutrino mass ordering.
    The current best-fit range for the Dirac phase $\delta$ is indicated by the white-shaded band.
    Parameter space regions incapable of producing the observed BAU value are shaded in blue.
    }
    \label{fig:BAUconstLowM}
\end{figure}

For a fixed triplet mass, the condition of successful leptogenesis allows us to deduce relations among low-energy parameters that are difficult to access experimentally, e.g., the Majorana phases $\sigma,\tau$.
In Fig.~\ref{fig:BAUconstLowM}, the triplet mass is fixed, $M_{\Delta}=1.2\cdot10^{13}$\ GeV for NO and $M_{\Delta}=0.64\cdot10^{13}$\ GeV for IO respectively, and the parameter space ensuring the correct value of the BAU is displayed in terms of pure low-energy parameters: the two CP-violating phases $\delta$ and $\sigma$ as well as the neutrino mass parameters $m_{ee}$ and $m_{\rm sm}$, respectively.
The remaining Majorana phase $\tau$ is again fixed to $\pi/2$.
For both mass orderings, distinct patterns can be identified and the general dependence on the Dirac and Majorana CP phases can be identified: a $2\pi$-periodicity for the Dirac CP phase $\delta$ and a $\pi$-periodicity for the Majorana phase $\sigma$.
For IO, the viable range of $m_{ee}$ and $m_{\rm sm}$ values is nearly independent of the Dirac CP phase $\delta$, while the contours strongly vary with $\delta$ in the NO case. 

\noindent This is a special feature for IO, since a given value for $m_{ee}$ can be directly cast into a constraint on the Majorana phase $\sigma$ and $\tau$ (the latter being fixed here), while guaranteeing at the same time the successful generation of the observed BAU with a triplet mass at the fixed value above.
Again, we recognize that IO generally allows for a larger effective Majorana masses than NO, 
with largest values obtained for $\sigma\sim 0,\pi$.
As mentioned before, the allowed region of type-II dominance explicitly depends on the CP phases.
Thus, we expect the situation to be changed when we allow the Majorana phase $\tau$ to vary as well.

Our assumptions of type-II dominance drives our model into parameter regions that are currently inaccessible by experiments that aim at measuring the effective electron neutrino mass $m_{\nu_{e}}$~\cite{Wolf:2008hf, Gastaldo:2017edk} or the effective Majorana mass $m_{ee}$, cf.\ Tab.~2 in Ref.~\cite{Dolinski:2019nrj}. 
On the other hand, the Dirac CP phase $\delta$ is expected to be tackled soon by future experiments~\cite{Abi:2020evt,Abe:2015zbg,Abe:2016ero}, which promises to result in further constraints on the parameter space of our model.

\section{Conclusions}
\label{sec:conclusions}


\noindent In this work, we investigated leptogenesis and low-energy CP violation in the context of a left-right-symmetric seesaw model.
While we left an explicit realization aside, the assumption of related model features in combination with type-II-dominated neutrino masses allowed us to establish a direct connection between high-energy and low-energy CP violation, which is the main focus of the present work, cf.\ Eq.~\eqref{eq:CPseparated}.
If we further approximate the Dirac mass matrix to have the same structure as the up-type quark mass matrix, which is motivated by $SO(10)$ GUTs, CP violation in our model can be fully parametrized by the phases of the PMNS lepton mixing matrix.
As a side effect, LH and RH neutrinos exhibit the same mass orderings, which are linked only by the ratio of corresponding scalar VEVs, cf.\ Eq.~\eqref{eq:linkedMasses}, and the available parameter space is only spanned by a few variables, most of them accessible at low energies.
The combination with a perturbativity bound on the lepton-triplet coupling $f$, cf.\ Eq.~\eqref{eq:perturbConstraint}, helps to reduce the parameter space further and gives our model predictive power in the context of our study.
The separation of LH and RH scales renders the RH neutrinos irrelevant for the investigation of leptogenesis, such that the baryon asymmetry of the Universe is solely generated by decays of one scalar $SU(2)_{L}$ triplet, cf.\ Fig.~\ref{fig:FeynmanDiagram}. 
Before that, the requirements for type-II dominance are investigated analytically and numerically as well as the consequences for the CP asymmetry $\epsilon$ generated by such decays are discussed, see Sec.~\ref{sec:MixedTypeIIDom} and \ref{sec:TypeIIDomCP}.
An updated scalar triplet leptogenesis framework based on density matrices including flavor and spectator corrections is introduced, cf.\ Eqs.~\eqref{eq:BEs}, while special care has been taken in the discussion of flavor effects within a scalar triplet framework, cf.\ Sec.~\ref{sec:LeptoFlavor}.
The discussed delay of flavor effects to due fast inverse triplet scatterings, cf.\ Eq.~\eqref{eq:newFlavorCond}, also affects the consideration of spectator processes.
Thus, the corresponding corrections have been refined and listed in Tab.~\ref{tab:spectatorCorr}.

It turned out that our choice of maximal, perturbatively allowed Yukawa couplings, cf.\ Eq.~\eqref{eq:perturbConstraint}, demands the triplet to decay dominantly into leptons.
This assumption in combination with the applied flavor criterion, cf.\ Eq.~\eqref{eq:newFlavorCond}, renders flavor effects to be of minor significance for leptogenesis.
In the following numerical investigation, the differences between our fully flavored treatment and a single-flavor approximation have been discussed and the impact of several ($\rm 2\rightarrow 2 $) washout processes is shown for a specific set of parameters, see Fig.~\ref{fig:AsymEvol} and Tab.~\ref{tab:WashoutComp}.
In addition, the influence of some parameters is discussed in more detail, while the robustness our results under varying experimental quantities has been checked, cf.\ Fig.~\ref{fig:CPPhases}.
Finally, we use the observed value of the baryon-to-photon ratio as an experimental constraint to establish further relations among the remaining model parameters.
In such way, we are able to link low-energy observables like the effective neutrino mass $m_{ee}$ to high-energy quantities such as the triplet mass $M_{\Delta}$, see Fig.~\ref{fig:BAUconstTripletMass}.
Complementary to this, relations among low-energy parameters can be deduced by assuming successful leptogenesis at a certain scale, cf.\ Fig.~\ref{fig:BAUconstLowM}.
Here, the full predictive power of our model becomes noticeable: 
Based on our assumptions, it is possible to shrink the generic parameter space of type-I and type-II seesaw models, and together with consistency bounds like perturbativity and the observed baryon-to-photon ratio $\eta_B$, the remaining parameter space collapses down to a handful of parameters ($m_{\rm sm}$, the mass ordering, $\delta$, $\sigma$, $\tau$, $M_{\Delta}$).
Knowledge of the four low-energy parameters would then allow us to infer (at least) bounds on the triplet mass $M_{\Delta}$ that lead to successful leptogenesis.
This is a remarkable result for a minimal scenario that manages to successfully explain neutrino oscillations as well as the baryon asymmetry of the Universe based on only a handful of parameters most of which will likely be measured or constrained further by experiments in the near future.

Of course, this minimal scenario relies on some strong assumptions and further studies have to show how to preserve this strong connection between high-energy and low-energy parameters when these assumptions are relaxed.
For example, open question are how to guarantee the model's predictive power if a smaller lepton-triplet coupling was used, or type-I and type-II seesaw scales are of the same order.
The first one will come along with an increasing impact of flavor effects since flavor transition bounds, cf.\ Eq.~\eqref{eq:newFlavorCond}, get weaker when the triplet does not only decay into leptons.
Here, a detailed treatment of flavored inverse triplet decays would be in order.
The second aspect would rely on a more general treatment taking into account all heavy particles involved, which includes further decay processes that contribute to the baryon asymmetry.
Further, the renormalization group running of all relevant quantities might be crucial for a detailed comparison with experimental measurements as well as more precise statements at highest energy.
When low-energy quantities are pinned further down by next-generation experiments, one will, within this framework, be able to establish further connections between low-energy observables and high-energy parameters, i.e.\ the triplet mass $M_{\Delta}$, rendering leptogenesis a predictive tool for additional low-energy parameters such as the remaining CP phases.


\section*{Acknowledgments}

\noindent T.\,R.\ is supported by the German Research Foundation (DFG) through the research training group GRK 1940 and the International Max Planck Research School for Precision Tests of Fundamental Symmetries.
W.\,R.\ is supported by the German Research Foundation (DFG) under grant number RO 2516/7-1 in the Heisenberg programme.
This project has received funding from the European Union's Horizon 2020 Research and Innovation Programme under grant agreement number 796961, "AxiBAU" (K.\,S.). 


\appendix


\section{Important formulas}
\label{app:AppA}


\noindent Let us summarize the ingredients that are necessary to solve the set of Boltzmann equations in Eqs.~\eqref{eq:BEs}. 
The Hubble parameter is given by
\begin{align}
\label{eq:Hubble}
H(z)=\sqrt{\frac{\pi^{2}}{3M_{\rm Pl}^{2}}\frac{g_{\rm SM}}{30}} \left(\frac{M_{\Delta}}{z}\right)^{2}\, ,
\end{align}
with $g_{\rm SM} = 106.75$ the number of degrees of freedom at high temperatures and the reduced Planck mass $M_{\rm Pl}= 2.435 \cdot 10^{18}$ GeV. 
The entropy density is given as 
\begin{align}
\label{eq:entropydensity}
s(z)=\frac{2\pi^{2} g_{\rm SM}^{*}}{45}\left(\frac{M_{\Delta}}{z}\right)^{3}\, \,
\end{align}
with $g_{\rm SM}^{*} = 106.75$.
Regarding number densities, we use the non-relativistic one for the description of triplet dynamics,
\begin{align}\label{eq:NeqMB}
    n^{\rm eq}_{\rm MB}(z,g)=\frac{g M_{\Delta}^{3}}{2\pi^{2}}
\frac{\mathcal{K}_{2}(z)}{z}\, ,
\end{align}
with $g$ representing the number of internal degrees of freedom and $\mathcal{K}_{n}(z)$ the modified Bessel function of the second kind.
The relativistic number densities are used especially for leptonic and scalar particle densities,
\begin{align}\label{eq:NeqBF}
    n^{\rm eq}(z,g)=\frac{\zeta(3) g}{\pi^{2}}\left(\frac{M_{\Delta}}{z}
    \right)^{3} \times 
    \begin{cases}
    1\, , & \text{for bosons}\, , \\
    \frac{3}{4}\,  & \text{for fermions}\, ,
    \end{cases}
\end{align}
where $g$ is the number of internal degrees of freedom of the particle. 
The equilibrium densities $Y^{\rm eq}$ follow from number densities $n^{\rm eq}$ by dividing them by the entropy density $s$; in particular, for scalar triplets this yields:
\begin{align}\label{eq:SDeq}
\Sigma^{\rm eq}_{\Delta}(z)=2\ \frac{n^{\rm eq}_{\rm MB}(z,3)}{s(z)}\, .
\end{align}
For fermions and bosons (lepton and Higgs doublets in our case), we apply $Y^{\rm eq}_{\rm L,H}(z) = n^{\rm eq}_{F,B}(z,\, 2) / s(z)$. 


\subsection*{Reaction densities}

\noindent In order to solve the Boltzmann equation given by Eqs.~\eqref{eq:BEs}, we need the reaction densities, which describe the number of reaction events per spacetime volume, that is, per spatial and temporal volume.
In the following, we list all quantities that are needed in the context of our investigation

The reaction density for decays and inverse decays is obtained via 
\begin{align}\label{eq:gammaD}
	\gamma_D (z) = \gamma\left(\Delta \leftrightarrow HH \right)
    \text{+}\sum_{\alpha,\beta} \gamma\left(\Delta \leftrightarrow 
    \overline{L}_{\alpha} \overline{L}_{\beta} \right)\text{+}\text{\,CP-conj.}
    = s(z)\Sigma^{\rm eq}_{\Delta}(z) \frac{\mathcal{K}_{1}(z)}{\mathcal{K}_{2}(z)}
    \Gamma_{\Delta}\, .
\end{align}

\begin{figure}
	\centering
	\includegraphics[width=0.95\textwidth]{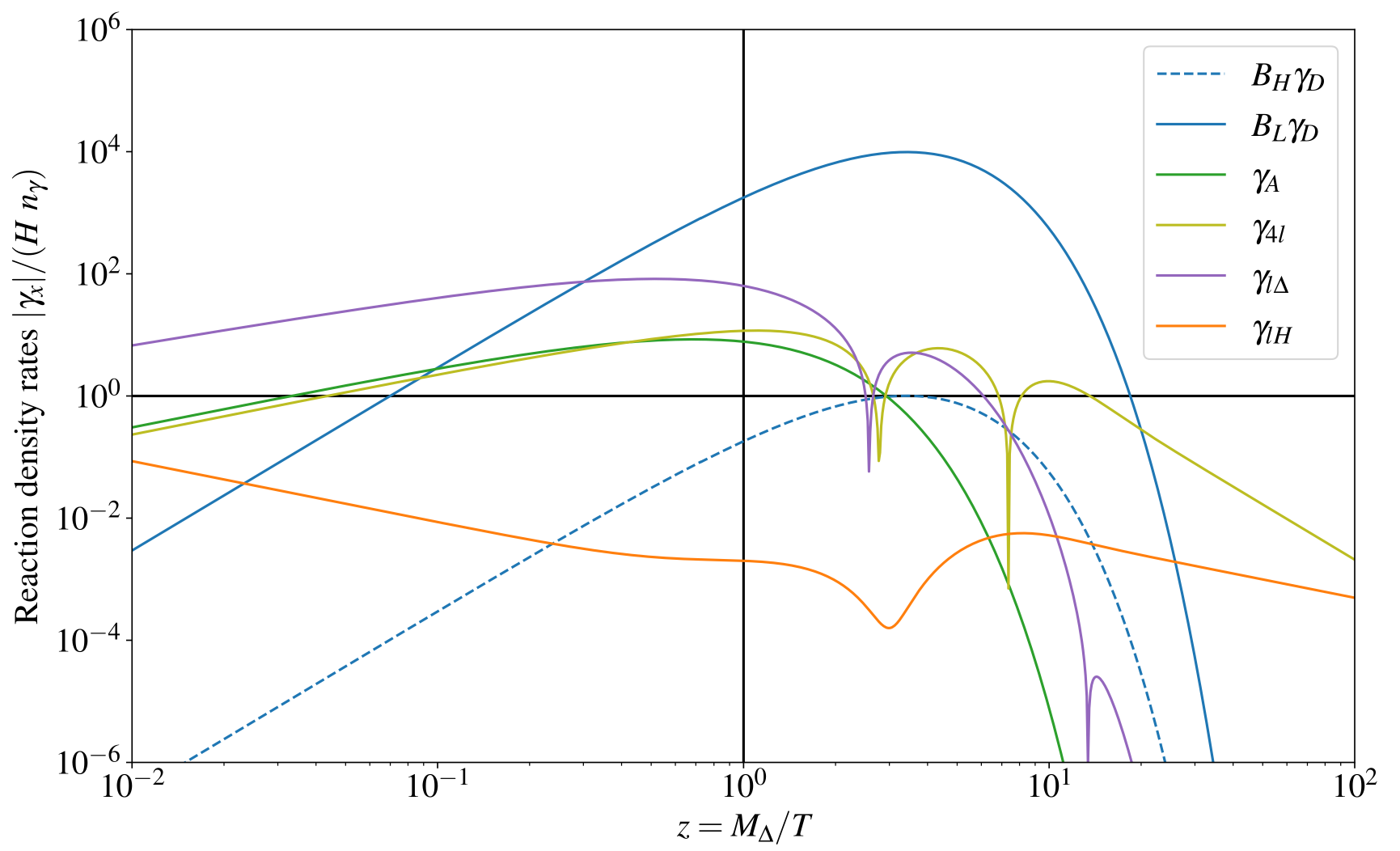}
	\caption{Reaction densities of decay and scattering processes applied in our framework with parameters according to Tab.~\ref{tab:fixedParameters}.
	The behavior for normal mass ordering is shown and all densities are normalized to $H n_{\gamma}$.}
	\label{fig:ReactionDensity}
\end{figure}

\noindent The scalar $SU(2)_{L}$ triplet undergoes gauge scatterings and other $\rm (2\rightarrow2)$ reactions.
The corresponding reaction densities are calculated via
\begin{align}
\begin{aligned}
    \gamma_{S}(z)&=\frac{M_{\Delta}^{4}}{64 z \pi^{4}}\int_{x_{\rm min}}^{\infty}
    dx\sqrt{x}\mathcal{K}_{1}(z\sqrt{x}) \hat{\sigma}_{S}(x)\, , 
\end{aligned}
\end{align}
with $x=s_{\rm min}/M_{\Delta}^{2}$ and the reduced cross section $\hat{\sigma}_{S}$ (summed over internal degrees of freedom of initial and final particles) of the corresponding reaction.
For example, the reduced cross section for gauge scatterings is given by~\cite{Hambye:2005tk,Hambye:2012fh}
\begin{equation}\label{eq:gammaA}
\begin{aligned}
    \hat{\sigma}_{A}  &= \frac{2}{72} \Big\{ (15 C_{1} - 3 C_{2} )r
    + (5 C_{2} - 11 C_{1} ) r^{3}\\ 
    &+ 3( r^{2} - 1) \big[ 2 C_{1}+ C_{2}( r^{2} - 1)\big] \log{\frac{1+r}{1-r}} \Big\} 
    +\left( \frac{50 g^{4} + 41 g_{Y}^{4}}{48}\right) r^{3/2}\, , 
\end{aligned}
\end{equation}
with $A=\sqrt{1-\frac{4}{x}}$, 
$C_{1} = 12 g^{4} + 3 g_{Y}^{4} + 12 g^{2} g_{Y}^{2}$, 
$C_{2}=6 g^{4} + 3 g_{Y}^{4} + 12 g^{2} g_{Y}^{2}$. 
Other reduced cross sections relevant for ($\rm 2\rightarrow 2$) washout processes can be found in the appendix of Ref.~\cite{Lavignac:2015gpa}. An illustration of reaction densities relevant for the following washout contributions can be seen in Fig.~\ref{fig:ReactionDensity} for NO with the parameters listed in Tab.~\ref{tab:fixedParameters}.

Now we list all washout terms that are used in our investigation, cf.\ Ref.~\cite{Lavignac:2015gpa}.
The most important ones are washout due to inverse lepton decays described by 
\begin{align}\label{eq:WD}
\begin{aligned}
	W^{D}=\frac{2B_{L}}{\lambda^{2}_{L}}
	\left[\frac{\Delta_{\Delta}}{\Sigma^{\rm eq}_{\Delta}}ff^{\dagger}+
	\frac{2f\Delta^{T}_{L}f^{\dagger}+ ff^{\dagger}\Delta_{L} + \Delta_{L}
	ff^{\dagger}}{4 Y^{\rm eq}_{L}} \right]\gamma_{D}\, ,
\end{aligned}
\end{align}
and washout induced by inverse Higgs decays ($HH\rightarrow \Delta, \overline{H}\overline{H}\rightarrow \overline{\Delta} $)  given by
\begin{align}\label{eq:WDHiggs}
	W^{H}=2B_{H}\left(\frac{\Delta_{H}}{Y_{H}^{\rm eq}}-
    \frac{\Delta_{\Delta}}{\Sigma^{\rm eq}_{\Delta}}\right)\gamma_{D}\, .
\end{align}
Further, we include several $(\Delta L=2)$ processes in our investigation.
For our model, the most important contribution comes from lepton-Higgs scatterings $W^{lH}$ ($L L\leftrightarrow\overline{H}\overline{H}$ and $LH \leftrightarrow \overline{L}\overline{H}$), given by
\begin{equation}\label{eq:WlH}
\resizebox{0.9\textwidth}{!}{$
    \begin{aligned}
    \frac{  W^{lH}}{2} &= \frac{\gamma^{\Delta}_{lH}}{\lambda^{2}_{L}}\left[ \frac{2f\Delta_{L}^{T} f^{\dagger}
    \text{+}ff^{\dagger}\Delta_{L}\text{+}\Delta_{L} ff^{\dagger}}{4Y^{\rm eq}_{L}}\text{+}\frac{\Delta_{H}}{Y^{\rm eq}_{H}}ff^{\dagger} \right]
    \text{+}\frac{\gamma^{\mathcal{H}}_{lH}}{\lambda^{2}_{\kappa}}\left[\frac{2\kappa\Delta_{L}^{T} 
    \kappa^{\dagger}\text{+}\kappa\kappa^{\dagger}\Delta_{L}\text{+}\Delta_{L} \kappa \kappa^{\dagger}}
    {4Y^{\rm eq}_{L}}\text{+}\frac{\Delta_{H}}{Y^{\rm eq}_{H}} \kappa\kappa^{\dagger}\right]
    \\
    & + \frac{\gamma^{\mathcal{I}}_{lH}}{{\rm Re}\left[{\rm Tr} \left[f \kappa^{\dagger}\right]\right]} 
    \left[\frac{2f\Delta_{L}^{T} \kappa^{\dagger}\text{+}\kappa\Delta_{L}^{T}f^{\dagger}
    \text{+}(f\kappa^{\dagger}\text{+}\kappa f^{\dagger}) \Delta_{L}\text{+}\Delta_{L} (f\kappa^{\dagger}\text{+}\kappa f^{\dagger})}{4Y^{\rm eq}_{L}}
    \text{+}\frac{\Delta_{H}}{Y^{\rm eq}_{H}} (f\kappa^{\dagger}\text{+}\kappa f^{\dagger})\right] ,
    \end{aligned}
    $}
\end{equation}
with $\kappa \simeq 2 v_{R} / v_{EW}^{2} m_{I}$ in the context of our model.
Further, contributions are coming from flavor-changing 2-lepton--2-lepton scatterings $W^{4l}$ ($L_{\alpha} L_{\beta}
\leftrightarrow L_{\gamma}L_{\delta}$ and $L_{\alpha} \overline{L}_{\gamma}\leftrightarrow
\overline{L}_{\beta}L_{\delta}$),
\begin{align}\label{eq:W4l}
    W^{4l} = \frac{2\gamma_{4l}}{\lambda^{4}_{L}Y^{\rm eq}_{L}} \left[ 
    \frac{\lambda^{2}_{L}}{4}(2f\Delta_{L}^{T} f^{\dagger} + f f^{\dagger} \Delta_{L} 
    + \Delta_{L} f f^{\dagger}) -{\rm Tr} (\Delta f f^{\dagger}) f f^{\dagger}\right] ,
\end{align}
and flavor-changing lepton-triplet scatterings ${W}^{l\Delta}$ ($L_{\alpha}\Delta
\leftrightarrow L_{\beta}\Delta$, $L_{\alpha}\overline{\Delta}\leftrightarrow
L_{\beta}\overline{\Delta}$ and $L_{\alpha}\overline{L}_{\beta}\leftrightarrow \Delta\overline{\Delta}$), 
\begin{align}\label{eq:WlDelta}
    W^{l\Delta}=\frac{\left[ff^{\dagger}ff^{\dagger}\Delta_{L} - 2ff^{\dagger}
	\Delta_{L} ff^{\dagger}	+\Delta_{L} ff^{\dagger}ff^{\dagger}\right]}{2Y^{\rm eq}_{L}
	{\rm Tr}\left[ff^{\dagger}ff^{\dagger}\right]}\gamma_{L\Delta}\, .
\end{align}


\section{Flavor and spectator corrections}\label{app:AppC}

\noindent Throughout this work we apply the following formula for the interaction rate of lepton Yukawa interactions~\cite{Weldon:1982bn,Cline:1993bd},
\begin{align}\label{eq:leptonYukawa}
    \Gamma_{f}(T) \simeq 5\cdot 10^{-3} y_{f}^{2}\,T \,, \quad \text{with } f=\tau, \mu\, .
\end{align}
Demanding it to be equal to the Hubble rates, $\Gamma\simeq H$, leads to the characteristic equilibrium temperatures $T^{\mu}\sim10^{9}$\ GeV and $T^{\tau}\sim10^{12}$\ GeV, respectively.

Further, we have introduced counter-terms in the Boltzmann equations~\eqref{eq:BEs} that take into account potential effects of the above processes.
Once they reach equilibrium, the following expressions drive the corresponding off-diagonal terms in $\Delta_{\alpha\beta}$ to zero,  
\begin{equation}\label{eq:YukawaCorr}
\begin{aligned}
	C_{\mu}(z, M_{\Delta})&= sig(z, z^{\mu}_{\rm dec}) s(z, M_{\Delta})
	\begin{pmatrix}
	0 & \Delta_{e\mu}& 0 \\ \Delta_{\mu e} & 0 & \Delta_{\mu\tau}\\ 0 & 
    \Delta_{\tau \mu} & 0 
	\end{pmatrix}, \\ \quad
	C_{\tau}(z, M_{\Delta})&= sig(z, z^{\tau}_{\rm dec}) s(z, M_{\Delta})
	\begin{pmatrix}
	0 & 0 & \Delta_{e \tau}  \\ 0 & 0 & \Delta_{\mu\tau}\\ \Delta_{\tau e} 
    & \Delta_{\tau \mu} & 0 
	\end{pmatrix} ,
\end{aligned}
\end{equation}
where the sigmoid function $sig(z,z^{x}_{\rm dec})$ activates the above counter-terms, at times $z^{x}_{\rm dec}$ when the lepton Yukawa interactions $x=\mu,\tau$ fall into equilibrium.\footnote{Conventionally, the evolution of dynamical quantities is described in terms of the dimensionless variable $z=M_{\Delta}/T$, where $T$ denotes the temperature of the Universe and $M_{\Delta}$ the mass of the decaying particle, here the scalar triplet. The advantage is that a lepton asymmetry is typically generated when the particle of interest becomes non-relativistic, thus, when $z\sim 1$. When carrying an index $x$, $z_{x}$ refers to a characteristic time, or equivalently, temperature $T_{x}$, when a certain reaction becomes relevant/irrelevant.}
The times $z^{x}_{\rm dec}$ are obtained by the modified flavor condition Eq.~\eqref{eq:newFlavorCond}. 
Note that the temperatures obtained from Eq.~\eqref{eq:leptonYukawa} are still necessary requirements. 

\subsection*{Weak flavor effects}

\begin{figure}[t]
    \centering
	\includegraphics[width=0.9\textwidth]{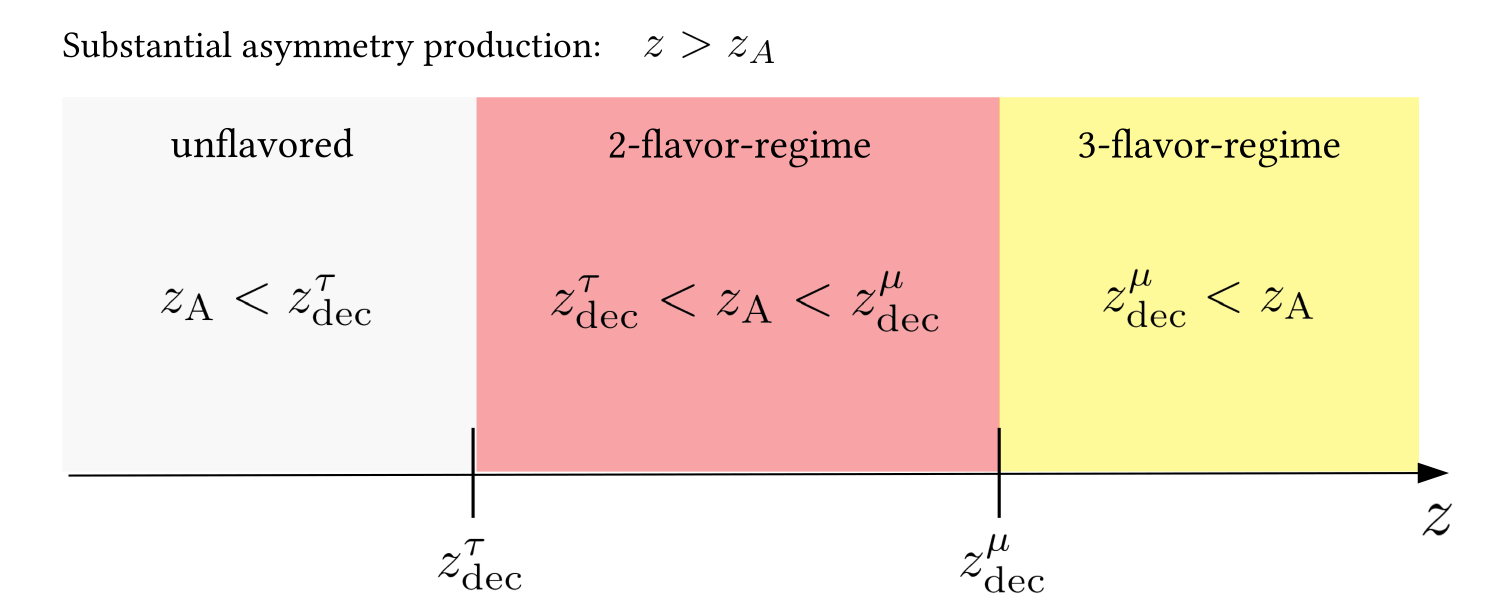}
	\caption{Schematic illustration of different flavor regimes in dependence of the time $z_{A} \propto 1/T$, when gauge scatterings become slower than triplet decays. 
	Within a flavored type-II framework, an asymmetry is efficiently produced for $z>z_{A}$ (necessary), while the explicit regime depends on the interplay between $z_{\rm dec}^{\tau,\mu}$ and $z_{A}$ (sufficient).} 
	\label{fig:FlavorRegimes}
\end{figure}

\noindent In Sec.~\ref{sec:LeptoFlavor}, we mentioned that fast gauge boson scattering, due to the triplet's weak charge, may influence the effects of appearing flavor decoherence.
From the viewpoint of these gauge interactions, sufficient asymmetry production is inhibited until the corresponding rate drops below the usual decay rate, $\Gamma_{A} < \Gamma_{D}$.
Hence, the time of effective BAU production is restricted to $z>z_{A}$, where $z_{A}$ is the time when gauge scatterings become slower than triplet decays.  
Taking into account lepton Yukawa interactions, this implies that, if $z^{x}_{\rm dec}\lesssim z_{A}$ for $x=\tau, \mu$, any asymmetry-generating interaction occurring at $z<z_{\rm dec}^{x}$ is of minor importance, and we can directly switch to the framework in which the corresponding flavor is already projected out.
By contrast, the symmetry production is dominated by the coherent regime if lepton Yukawa interactions  reach equilibrium long after gauge interactions, $z_{\rm dec}\gg z_{A}$. 
In addition, this is supported by the fact that for $z>z_{\rm dec}$ the triplet abundance is already expected to be Boltzmann suppressed, which directly affects the asymmetry production.
Therefore, a necessary condition for the generation of an asymmetry is $z>z_{A}$, while for the transition of flavor regimes, the condition $z^{\tau,\mu}_{\rm dec}\lesssim z_{A}$ is sufficient. 
The whole picture including both relevant lepton Yukawa interactions is illustrated in Fig.~\ref{fig:FlavorRegimes}.
In general, the transition conditions induced by gauge scattering are assumed to be weaker than the ones from Eq.~\eqref{eq:newFlavorCond}.
A detailed investigation of this condition has been performed in Ref.~\cite{Sierra:2014tqa}.


\subsection*{Modified spectator corrections}

\noindent The modification of flavor transitions according to Eq.~\eqref{eq:newFlavorCond} has consequences for  the spectator corrections as well. 
Although the equilibrium temperature of the individual processes is independent of any triplet interaction, from the viewpoint of  successful creation of a lepton asymmetry, triplet interactions and their timescale become crucial.
As stated in Sec.~\ref{sec:LeptoFlavor}, triplet inverse decays may proceed faster than lepton Yukawa interactions, such that the latter effectively remain out of equilibrium.
This treatment has to be considered in the formulation of chemical equilibrium conditions and leads to modifications in the spectator corrections. 
This modified spectator corrections (in the convention of Ref.~\cite{Lavignac:2015gpa}) are listed in Tab.~\ref{tab:spectatorCorr}.


\begin{sidewaystable}
\resizebox{0.95\textwidth}{!}{
\centering
\begin{tabular}{c|c|c|c|c}
$T$ [GeV] & In equilibrium & Flavor(s) & Global symmetries of epoch & spectator correction\\
\hline 
&&&&\\
$\gtrsim 10^{15}$ & Hyp.   & 1 & $ U(1)_{Y}\times U(1)_{ B}\times U(1)_{e_{R}}\times U(1)_{PQ} 
\times SU(3)_{Q} \times SU(3)_{u} \times SU(3)_{d} \times SU(3)_{e}$ & 
$\rm \begin{aligned}
\left( \Delta_{L}\right)_{\alpha\beta} = -\Delta_{\alpha\beta}\, , \quad 
\Delta_{H} =-\text{Tr}(\Delta_{_{\alpha\beta}}) - 2\Delta_{\Delta} 
\end{aligned}$
\\
&&&&\\
\hline
&&&&\\
$[10^{13}, 10^{15}]$ & Hyp., $t$ & 1 & $ U(1)_{Y}\times U(1)_{B}\times U(1)_{PQ} \times U(1)_{e_{R}}
\times SU(2)_{Q} \times SU(2)_{u} \times SU(3)_{d} \times SU(3)_{e}$ & 
$\rm \begin{aligned}
\left( \Delta_{L}\right)_{\alpha\beta} = -\Delta_{\alpha\beta}\, , \quad 
\Delta_{H} =- \frac{2}{3}\text{Tr}(\Delta_{_{\alpha\beta}}) - \frac{4}{3}\Delta_{\Delta} 
\end{aligned}$
\\
&&&&\\
\hline
%
&&&&\\
$[10^{12}, 10^{13}]$ &  Hyp., Sphal., $t$ & 1 & $ U(1)_{Y}\times U(1)_{B}\times U(1)_{e_{R}} 
\times SU(2)_{Q} \times SU(2)_{u} \times SU(3)_{d} \times SU(3)_{e}$ & 
$\rm \begin{aligned}
\left( \Delta_{L}\right)_{\alpha\beta} = -\Delta_{\alpha\beta}\, , \quad
\Delta_{H} =- \frac{14}{23} \text{Tr}(\Delta_{_{\alpha\beta}}) - \frac{28}{23}\Delta_{\Delta} 
\end{aligned}$
\\
&&&&\\
\hline
$[10^{9}, 10^{12}]$: & & & & \\ 
%
$ \quad [T^{\tau}_{\rm dec}, 10^{12}]$ & Hyp., Sphal.,$t,b,c$, & 1 & 
$ U(1)_{Y}\times U(1)_{Q}\times U(1)_{u} \times SU(2)_{d} \times SU(3)_{e}$ & 
$\rm \begin{aligned}
\left( \Delta_{L}\right)_{\alpha\beta} = -\frac{3}{5} \Delta_{\alpha\beta}\, , \quad
\Delta_{H} = -\frac{4}{13}\text{Tr}(\Delta_{\alpha\beta})- \frac{8}{13} \Delta_{\Delta }
\end{aligned}$
\\
&&&&\\
$ \quad [10^{9}, T^{\tau}_{\rm dec}]$ & Hyp., Sphal.,$t,b,c, \tau$ & 2 & 
$ U(1)_{Y}\times U(1)_{Q}\times U(1)_{u} \times SU(2)_{d} \times SU(2)_{e}$ & 
$\rm \begin{aligned}
\left( \Delta_{L}\right)_{\alpha\beta} & =
\begin{pmatrix}
\multicolumn{2}{c}{ \multirow{2}{*}{ $\left(\frac{86}{589} \text{Tr}\left(\Delta_{ij}\right) + \frac{60}{589} \Delta_{33} + \frac{8}{589}\Delta_{\Delta}\right) \delta_{ij}-\Delta_{ij}$ }}  & 0 \\
& & 0\\
0 & 0 & \frac{30}{589} \text{Tr}\left( \Delta_{ij} \right) -\frac{390}{589}\Delta_{33} - \frac{52}{589}\Delta_{\Delta}
\end{pmatrix} \\
\Delta_{H} & =-\frac{164}{589} \text{Tr}(\Delta_{ij}) - \frac{224}{589}\Delta_{\tau} - \frac{344}{589}\Delta_{\Delta} 
\end{aligned}$\\
&&&&\\
\hline
$[10^{5}, 10^{9}]$: & & & & \\
%
$ \quad [T^{\tau}_{\rm dec}, 10^{9}]$ & Hyp., Sphal.,$t,b,c,s$ & 1 & 
$ U(1)_{Y}\times U(1)_{Q}\times U(1)_{u} \times SU(2)_{d} \times SU(3)_{e}$ & 
$\rm \begin{aligned}
\left( \Delta_{L}\right)_{\alpha\beta} = -\frac{3}{5} \Delta_{\alpha\beta}\, , \quad
\Delta_{H} = -\frac{1}{4}\text{Tr}(\Delta_{\alpha\beta})- \frac{1}{2} \Delta_{\Delta }
\end{aligned}$
\\
&&&&\\
$ \quad [T^{\mu}_{\rm dec}, T^{\tau}_{\rm dec}]$ & Hyp., Sphal.,$t,b,c,s, \tau$ & 2 & 
$U(1)_{Y}\times U(1)_{Q}\times U(1)_{u} \times SU(2)_{d} \times SU(2)_{e}$ & 
$\rm \begin{aligned}
\left( \Delta_{L}\right)_{\alpha\beta} & =
\begin{pmatrix}
\multicolumn{2}{c}{ \multirow{2}{*}{ $\left(\frac{52}{359} \text{Tr}\left(\Delta_{ij}\right) + \frac{36}{359} \Delta_{33} + \frac{4}{359}\Delta_{\Delta}\right) \delta_{ij}-\Delta_{ij}$ }}  & 0 \\
& & 0\\
0 & 0 & \frac{21}{359} \text{Tr}\left( \Delta_{ij} \right) -\frac{234}{359}\Delta_{33} - \frac{26}{359}\Delta_{\Delta}
\end{pmatrix} \\
\Delta_{H} & =-\frac{82}{359} \text{Tr}(\Delta_{ij}) - \frac{112}{359}\Delta_{\tau} - \frac{172}{359}\Delta_{\Delta}
\end{aligned}$
\\
&&&&\\
$ \quad [10^{5}, T^{\mu}_{\rm dec}]$ & Hyp., Sphal.,$t,b,c,s,\tau,\mu$ & 3 & 
$ U(1)_{Y}\times U(1)_{Q}\times U(1)_{u} \times SU(2)_{d} \times U(1)_{e}$ & 
$\rm \begin{aligned}
\left( \Delta_{L}\right)_{\alpha\beta} &= 
\text{Diag}\bigg[ -\frac{151}{179}\Delta_{11} + \frac{20}{179}\Delta_{22} + \frac{20}{179}\Delta_{33} + \frac{4}{179}\Delta_{\Delta}\, , \frac{25}{358}\Delta_{11} -\frac{344}{537}\Delta_{22} + \frac{14}{537}\Delta_{33} - \frac{11}{179}\Delta_{\Delta}\, ,\\
& \quad\quad\quad\quad \frac{25}{358}\Delta_{11} +\frac{14}{537}\Delta_{22} - \frac{344}{537}\Delta_{33} - \frac{11}{179}\Delta_{\Delta} \bigg]\\ 
\Delta_{H} & =-\frac{37}{179} \Delta_{11}-\frac{52}{179} \Delta_{22} - \frac{52}{179}\Delta_{33}- \frac{82}{179}\Delta_{\Delta} 
\end{aligned}$
\\
&&&&\\
\hline
$\lesssim 10^{5}$: & & & &\\
$ \quad [T^{\tau}_{\rm dec}, 10^{5}]$ & Hyp., Sphal.,$t,b,c,s,u,d$ & 1 & 
$ U(1)_{Y}\times SU(3)_{e}$ &
$\rm \begin{aligned}
\left( \Delta_{L}\right)_{\alpha\beta} = -\frac{3}{5} \Delta_{\alpha\beta}\, , \quad
\Delta_{H} = -\frac{2}{11}\text{Tr}(\Delta_{\alpha\beta}) - \frac{4}{11} \Delta_{\Delta }
\end{aligned}$
\\
&&&&\\
$ \quad [T^{\mu}_{\rm dec}, T^{\tau}_{\rm dec}]$ & Hyp., Sphal.,$t,b,c,s,u,d,\tau$ & 2 & 
$ U(1)_{Y}\times SU(2)_{e}$ & 
$\rm \begin{aligned}
\left( \Delta_{L}\right)_{\alpha\beta} & =
\begin{pmatrix}
\multicolumn{2}{c}{ \multirow{2}{*}{ $\left(\frac{35}{244} \text{Tr}\left(\Delta_{ij}\right) + \frac{6}{61} \Delta_{33} + \frac{1}{122}\Delta_{\Delta}\right) \delta_{ij}-\Delta_{ij}$ }}  & 0 \\
& & 0\\
0 & 0 & \frac{33}{488} \text{Tr}\left( \Delta_{ij} \right) -\frac{39}{61}\Delta_{33} - \frac{13}{244}\Delta_{\Delta}
\end{pmatrix} \\
\Delta_{H} & =-\frac{41}{244} \text{Tr}(\Delta_{ij}) - \frac{14}{61}\Delta_{\tau} - \frac{43}{122}\Delta_{\Delta} 
\end{aligned}$
\\
 &&&&\\
$\quad [T^{e}_{\rm dec}, T^{\mu}_{\rm dec}]$ & Hyp., Sphal.,$t,b,c,s,u,d,\tau,\mu$ & 3 & 
$ U(1)_{Y}\times U(1)_{e}$ &
$\rm \begin{aligned}
\left( \Delta_{L}\right)_{\alpha\beta} &= 
\text{Diag}\bigg[ -\frac{11}{13}\Delta_{11} + \frac{4}{37}\Delta_{22} + \frac{4}{37}\Delta_{33} + \frac{8}{481}\Delta_{\Delta}\, , \frac{1}{13}\Delta_{11} -\frac{70}{111}\Delta_{22} + \frac{4}{111}\Delta_{33} - \frac{22}{481}\Delta_{\Delta}\, ,\\
& \quad\quad\quad\quad \frac{1}{13}\Delta_{11} +\frac{4}{111}\Delta_{22} - \frac{70}{111}\Delta_{33} - \frac{22}{481}\Delta_{\Delta} \bigg]\\ 
\Delta_{H} & =-\frac{2}{13} \Delta_{11}-\frac{8}{37} \Delta_{22} - \frac{8}{37}\Delta_{33}- \frac{164}{481}\Delta_{\Delta} 
\end{aligned}$
\\
&&&&\\
$\lesssim T^{e}_{\rm dec}$ & Hyp., Sphal.,$t,b,c,s,\tau,\mu,e$ & 3 & 
$ U(1)_{Y}$ &
$\rm \begin{aligned}
\left( \Delta_{L}\right)_{\alpha\beta} &= 
\text{Diag}\bigg[ -\frac{422}{711}\Delta_{11} + \frac{32}{711}\Delta_{22} + \frac{32}{711}\Delta_{33} - \frac{3}{79}\Delta_{\Delta}\, , \frac{32}{711}\Delta_{11} -\frac{442}{711}\Delta_{22} + \frac{32}{711}\Delta_{33} - \frac{3}{79}\Delta_{\Delta}\, ,\\
& \quad\quad\quad\quad \frac{32}{711}\Delta_{11} +\frac{32}{711}\Delta_{22} - \frac{442}{711}\Delta_{33} - \frac{3}{79}\Delta_{\Delta} \bigg]\\ 
\Delta_{H} & =-\frac{16}{79} \Delta_{11}-\frac{16}{79} \Delta_{22} - \frac{16}{79}\Delta_{33}- \frac{26}{79}\Delta_{\Delta} 
\end{aligned}
$
\\
&&&&\\
\end{tabular}
}
\caption{\small Temperature ranges in the early Universe with corresponding reactions 
in thermal equilibrium.
The third and fourth column show the flavor regime that has to be applied and
the global symmetries of the early Universe effective Lagrangian.
In the fifth column, the individual spectator corrections are listed for the 
corresponding temperature regimes.
These corrections incorporate the modifications due to fast inverse triplet decays, 
cf.\ Eq.~\eqref{eq:newFlavorCond}. The indices $i,j$ represent the flavor subspace 
orthogonal to the $\tau$-direction, thus $i,j=1,2$. }
\label{tab:spectatorCorr}
\end{sidewaystable}



\section{Explicit expressions for type-II-dominated CP asymmetry}\label{app:AppB}

\begin{figure}[t]
    \centering
    \includegraphics[width=\textwidth]{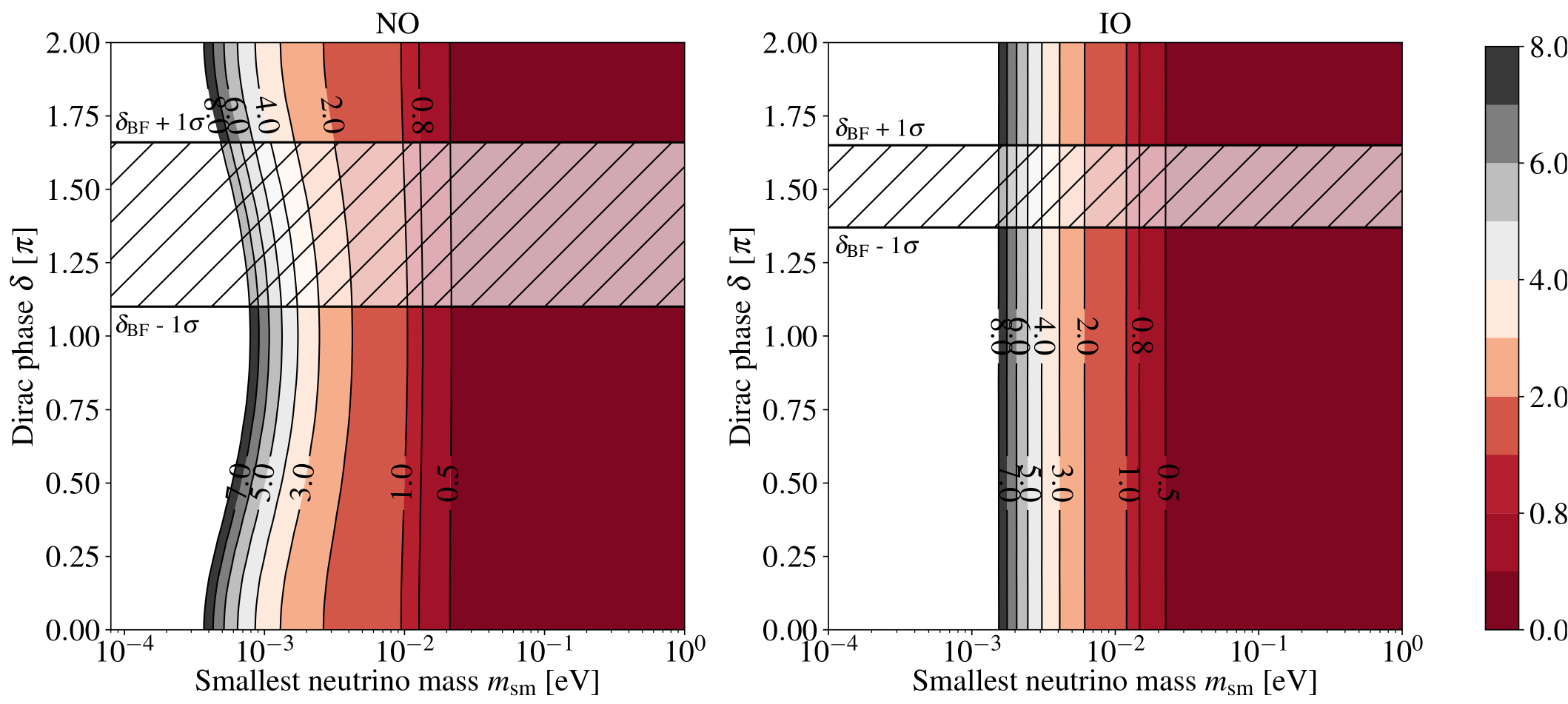}
    \caption{Contour plots of $A_{\rm max}$ for normal (left) and inverted (right) mass ordering, respectively.
    Horizontal lines indicate the $1\sigma$ range of the Dirac CP phase from a recent global fit~\cite{Capozzi:2020qhw}.}
    \label{fig:Amax}
\end{figure}

\noindent In Section \ref{sec:TypeIIDomCP} we have seen that, in our framework, the CP asymmetry \eqref{eq:CPasym} can be separated into quantities that only contain low-energy  and high-energy parameters, respectively.
These are obtained by assuming type-II dominance, which simplifies the type-I and type-II light neutrino mass contributions. 
Further, application of the perturbativity condition in Eq.~\eqref{eq:perturbConstraint} leads to the final separation of parameter space.
The low-energy function $A$ from Eq.~\eqref{eq:CPseparated} depends on the lightest active-neutrino mass $m_{\rm sm}$ as well as on the CP-violating phases $\delta$, $\sigma$ and $\tau$, such that it can, in principle, obtain arbitrarily small absolute values. 
In the following, we will be interested in its maximally possible values by maximizing it over the Majorana CP phases $\sigma$ and $\tau$,  
\begin{align}
A_{\rm max}(m_{\rm sm}, \delta)= \max_{\sigma,\tau} \left|A(m_{\rm sm}, \delta, \sigma, \tau)\right| .
\end{align}
As evident from Fig.~\ref{fig:Amax}, $A_{\rm max}$ is always of $\mathcal{O}(1)$ as long as $m_{\rm sm} \sim 0.01$\ eV.
In this part of parameter space, the parameter $B$, cf.\ Eq.~\eqref{eq:CPseparated}, can therefore be regarded as a good estimate for the maximal possible CP parameter $|\epsilon_{\Delta}|$.


\subsection*{Benchmark value for scalar triplet mass}
 

\noindent Setting $\lambda_{H}=1$ results in an important benchmark value for the scalar triplet mass, namely 
\begin{align}\label{eq:benchmarkMass}
M^{0}_{\Delta}\equiv\frac{v^{2}_{\rm ew}}{\overline{m}}\simeq 6 \cdot 10^{14}\ \text{GeV} 
\left(\frac{0.05\ \text{eV}}{\overline{m}}\right) .
\end{align}
This mass scale allows to distinguish between two different regimes regarding $B$ in Eq.~\eqref{eq:CPseparated},
\begin{equation}\label{eq:Bparam}
\resizebox{0.9\textwidth}{!}{$
B \simeq
\begin{cases} \displaystyle 
\frac{\overline{m} M_{\Delta}}
{ 8\pi \, v^{2}_{\rm ew}\, v_{R}}  \simeq 
3 \cdot 10^{-9} \left( \frac{\overline{m}}{0.05\ 
\text{eV}} \right) \left( \frac{10^{16}\ \text{GeV}}{v_{R}} \right) \left( 
\frac{M_{\Delta}}{10^{12}\ \text{GeV}} \right)^{2} 
& \mbox{ for }  M_{\Delta}\ll M^{0}_{\Delta}\, ,
\\ \displaystyle
\\ \displaystyle
\frac{v^{2}_{\rm ew}}{ 8 \pi\, \overline{m} \,v_{R}} \ \ \simeq
5 \cdot 10^{-3} \left( 
\frac{0.05\ \text{eV}}{\overline{m}} \right) \left(\frac{10^{16}\ \text{GeV}}{v_{R}}
\right)
& \mbox{ for }  M_{\Delta}\gg M^{0}_{\Delta}\, .
\end{cases}$}
\end{equation}
Thus, the function $B$ saturates at a value of $\mathcal{O}(10^{-3})$ once $M_{\Delta}$ is larger than $M_{\Delta}^{0}$. 


\subsection*{Explicit Dependence on CP phases}


\noindent Expression \eqref{eq:analyticA} is found by introducing the auxiliary parameters $R=\delta m^{2}/|\Delta m^{2}|$ and $\eta=m_{\rm sm}^{2}/|\Delta m^{2}|$, while setting $\theta_{23}\rightarrow\frac{\pi}{4}$ and $\sin^{2} \theta_{12} \rightarrow\frac{1}{3}$. 
Linearization in terms of magnitude, in particular $R\simeq\frac{1}{30}$ and $\theta_{13}^{2}\simeq \frac{1}{49}$ leads to the compact expression shown in Sec.~\ref{sec:TypeIIDomCP}, where the
following terms have been introduced
\begin{equation}\label{eq:appS}
    \begin{aligned}
    S_{1} &= \frac{1}{576} \big[16 \sin2 \sigma+ 16\sin4 \sigma +36 \sin 4 \tau \big]\\
    & -\frac{\sqrt{2}}{36} \theta_{13}\big[\sin (\delta +2 \sigma )-2 \sin (\delta +4 \sigma )+\sin\delta\big] \\
   & +\frac{\theta_{13}^{2}}{24} \big[2\sin2\delta -\sin 2( \delta + \sigma )+2 \sin 2( \delta +2 \sigma )   - 3 \sin 4\tau \big]\, ,\\
   S_{2} &= \sqrt{2}  \theta_{13} \big[\sin (\delta +2 (\sigma +\tau)) -\sin (\delta +2 \tau ) \big]\\
    & +  \frac{\theta_{13}^{2}}{2} \big[\sin 2( \delta + \sigma + \tau )+2 \sin 2( \delta +\tau )-2 \sin 2( \sigma + \tau )-\sin 2 \tau \big]\, ,\\
    T&=4(1+\eta)(1+2\eta)\, .
\end{aligned}
\end{equation}
Note that the full dependence on the CP phase $\delta$ has been cast into $S_{1,2}$.

\noindent More simplified formulas can be obtained by also taking into account $\eta$, while the resulting expressions depend on the value of the smallest neutrino mass, $m_{1}$ for NO and $m_{3}$ for IO respectively.
Independent of the light mass ordering, $\eta$ is approximately
\begin{align}
    \eta\simeq\begin{cases}
    R\, , & \text{for } m_{\rm sm}=10^{-2}\,\text{eV}\, , \\
    0 \, , & \text{for } m_{\rm sm}=10^{-4}\,\text{eV}\, .
    \end{cases}
\end{align}
In the context of leptogenesis, it must not be zero due to arising divergences in $A$, cf.\ Eq.~\eqref{eq:analyticA}.
For this reason, we only state the leading divergent terms for $m_{\rm sm}\sim 10^{-4}$\, eV. 
Furthermore, we give terms for specific CP phase configurations to highlight the characteristic dependence of Dirac and Majorana phases:
$\epsilon \propto \sin \delta$ in leading order $\theta_{13}$ for $\delta$ and $\epsilon \propto \sin 2\alpha$ and higher periodicities for $\alpha=\sigma,\tau$.


\paragraph{Normal ordering:} $m_{1}=10^{-2}\ \text{eV}$
\small{
\begin{align}
\begin{split}\label{eq:ANOhigh_gen}
    576\cdot A(m_{1}{}&=10^{-2}\ \text{eV}) =\widetilde{A_{1}} + \frac{6\theta_{13}^2}{\sqrt{R}} \bigg[\sqrt{2} (\sin 2 (\delta +\sigma +\tau )-2 \sin 2 (\sigma +\tau ))\\
    &+4 \sin 2 (\delta +\tau ) +\sqrt{R} \left((16 \cos 2 \sigma -3 \sqrt{2}) \sin2 (\delta +\sigma )-12 \sin4 \tau \right)-2 \sin 2 \tau \bigg]\\&
   + {2\theta_{13}} \bigg[\frac{12}{\sqrt{R}} \left(\sin (\delta +2
   (\sigma +\tau ))-\sqrt{2} \sin (\delta +2 \tau )\right)+33 \sqrt{R} \sin (\delta +2 (\sigma +\tau ))\\
   &+4 \left(-3 \sin (\delta +2 \sigma )+4 \sqrt{2} \sin (\delta +4
   \sigma )-2 \sqrt{2} \sin\delta \right)-21 \sqrt{2R} \sin (\delta +2 \tau )\bigg]\, , 
\end{split}\\
\begin{split}\label{eq:ANOhigh_st}
    576 \cdot A(m_{1}{}&=10^{-2}\ \text{eV}| \sigma,\tau=0) = 6\theta_{13}^2 \bigg[16-3 \sqrt{2}+\frac{\sqrt{2}+4}{\sqrt{R}}\bigg] \sin 2 \delta \\
     &+2\theta_{13}\bigg[\left(33-21 \sqrt{2}\right)\sqrt{R} +4 \left(2 \sqrt{2}-3\right) - \frac{12}{\sqrt{R}} \left(\sqrt{2}-1\right)\bigg] \sin\delta \, ,
\end{split}\\
\begin{split}
    576 \cdot A(m_{1}&=10^{-2}\ \text{eV}|\delta= \pi)  = \widetilde{A}_{1} + 6\theta_{13}^{2} \bigg[-3 \sqrt{2} \sin 2\sigma + 8 \sin 4\sigma -12 \sin 4\tau \\&- \frac{\sqrt{2}}{\sqrt{R}} \sin 2(\sigma +\tau ) + \frac{2}{\sqrt{R}} \sin 2\tau \bigg] + 2\theta_{13} \bigg[12 \sin 2\sigma - 16 \sqrt{2} \sin 4\sigma\\& +\frac{3 \sqrt{2} (7 R+4)}{\sqrt{R}} \sin 2\tau -\frac{3 (11 R+4)}{\sqrt{R}} \sin 2(\sigma + \tau) \bigg]\, ,
\end{split}\\
\begin{split}\label{eq:ANOhigh_d}
    576 \cdot A(m_{1}&=10^{-2}\ \text{eV}|\delta= 3\frac{\pi}{2})=\widetilde{A}_{1} +6\theta_{13}^{2} \bigg[-\frac{3}{\sqrt{R}} \left(\sqrt{2} \sin 2 (\sigma +\tau ) +2 \sin 2\tau \right)\\& +3 \sqrt{2} \sin 2\sigma  -8 \sin 4\sigma -12 \sin 4\tau \bigg] + 2\theta_{13} \bigg[12 \cos 2\sigma + 8 \sqrt{2} (1-2 \cos 4\sigma)\\& + 3 \sqrt{R} \left(7 \sqrt{2} \cos 2\tau  -11 \cos 2(\sigma +\tau ) \right) +\frac{12}{\sqrt{R}} \left(\sqrt{2} \cos 2\tau -\cos 2(\sigma +\tau) \right)\bigg]\, ,
\end{split}
\end{align}
}
with $\widetilde{A}_{1}=\bigg[\frac{3 \sqrt{2} (11
   R+4)}{\sqrt{R}} \sin 2(\sigma +\tau )+ \frac{3(7 R+4)}{\sqrt{R}} \sin 2\tau  +12 \sqrt{2} \sin 2\sigma +16 \sin 4\sigma +36 \sin 4\tau \bigg]$.


\paragraph{Normal ordering:} $m_{1}\sim10^{-4}\ \text{eV}$
\small{
\begin{align}
\begin{split}\label{eq:ANOlow_gen}
    576 \cdot A(m_{1}&\sim 10^{-4}\ \text{eV}) \simeq \widetilde{A}_{2} + 12\frac{\theta_{13}^2}{\sqrt{\eta }} \bigg[2 \sqrt{\eta } ((4 \cos 2\sigma - 1) \sin 2 (\delta +\sigma ) - 3 \sin 4\tau)\\& + 3 (3 \eta +2) \sin\delta \cos\sigma \cos (\delta +\sigma +2 \tau )
    - (3 \eta +2) \cos\delta \sin\sigma \cos (\delta +\sigma +2 \tau ) \bigg]\\
   &+16\frac{\theta_{13} \sin \sigma}{\sqrt{2\eta }} \bigg[3 (3 \eta +2) \cos (\delta +\sigma +2 \tau ) + 4 \sqrt{\eta } (\cos (\delta +\sigma )+2 \cos (\delta +3 \sigma ))\bigg]\, ,
\end{split}\\
\begin{split}\label{eq:ANOlow_st}
    576 \cdot A(m_{1}&\sim 10^{-4}\ \text{eV}|\sigma,\tau=0)\simeq 18\theta_{13}^2 \bigg[\frac{2}{\sqrt{\eta }}+ 4+ 3 \sqrt{\eta} \bigg]  \sin 2\delta\, ,
\end{split}\\
\begin{split}
    576 \cdot A(m_{1}&\sim 10^{-4}\ \text{eV}|\delta=\pi) \simeq \widetilde{A}_{2} +6 \theta_{13}^2 \bigg[3 \sqrt{\eta } \sin 2\tau - 4 \sin 2\sigma + 8 \sin 4\sigma - 12 \sin 4\tau \\&
    - 3 \sqrt{\eta}  \sin 2(\sigma +\tau ) - \frac{4}{\sqrt{\eta}} \sin \sigma \cos (\sigma +2 \tau)\bigg]\\&
   -16\frac{\theta_{13} }{\sqrt{2\eta }} \sin \sigma\bigg[3 (3 \eta +2) \cos (\sigma +2 \tau )+4 \sqrt{\eta } (\cos \sigma + 2 \cos 3\sigma)\bigg]
\end{split}
\end{align}
\begin{align}
\begin{split}\label{eq:ANOlow_d}
    576 \cdot A(m_{1}&\sim 10^{-4}\ \text{eV}|\delta=3\frac{\pi}{2}) \simeq \widetilde{A}_{2} + 6\theta_{13}^2 \bigg[4 \sin 2\sigma -8 \sin 4\sigma\\&
    -\frac{3}{\sqrt{\eta}} \bigg((3 \eta +2) \sin 2(\sigma +\tau )
    +(3 \eta +2) \sin 2\tau  +4 \sqrt{\eta } \sin 4\tau\bigg) 
    \bigg] \\&
   +16\frac{\theta_{13}}{ \sqrt{2\eta }} \sin \sigma \bigg[3 (3 \eta +2) \sin (\sigma +2 \tau )+4 \sqrt{\eta } (\sin\sigma +2 \sin 3\sigma )\bigg] \, ,
    \end{split}
\end{align}
}
with $\widetilde{A}_{2}=\bigg[ 16\sin 2\sigma +16\sin 4\sigma + 36 \sin 4\tau  + \frac{3 (8 \eta  (3 \eta +2)+((8-9 \eta ) \eta -8) R)}{2\sqrt{\eta^3}} \sin 2(\sigma +\tau ) + \frac{3 (12 \eta +(2-3 \eta ) R+8)}{2\sqrt{\eta}} \sin 2\tau \bigg]$\, .


\paragraph{Inverted ordering:} $m_{3}=10^{-2}\ \text{eV}$
\small{
\begin{align}
\begin{split}\label{eq:AIOhigh_gen}
    576 \cdot A(m_{3}&=10^{-2}\ \text{eV}) =\widetilde{A}_{3} + \frac{\theta_{13}^2}{24} \bigg[\frac{\cos (\delta +\sigma +2 \tau )}{\sqrt{R}}(2 \sin (\delta -\sigma )+\sin (\delta +\sigma ))\\& -3 \sin 4\tau +(4 \cos 2\sigma -1) \sin2 (\delta +\sigma)\bigg] +\frac{4\theta_{13}}{\sqrt{2R}} \bigg[24 \sin\sigma \cos (\delta +\sigma +2 \tau )\\& -15 R \sin (\delta +2 \tau ) \\
    &+21 R \sin (\delta +2 (\sigma +\tau )) +16 \sqrt{R} \sin \sigma (\cos (\delta +\sigma )+2 \cos (\delta +3 \sigma ))\bigg]\, ,
\end{split}\\
\begin{split}\label{eq:AIOhigh_st}
    576\cdot A(m_{3}&=10^{-2}\ \text{eV}|\sigma,\tau=0)= 36 \theta_{13}^2 \left(\frac{1}{\sqrt{R}} + 2\right)\sin 2\delta + 24 \theta_{13} \sqrt{\frac{R}{2}}  \sin \delta\, ,
\end{split}\\
\begin{split}
    576\cdot A(m_{3}&=10^{-2}\ \text{eV}|\delta=\pi)=\widetilde{A}_{3} + 24 \theta_{13}^2 \bigg[-\frac{1}{\sqrt{R}}\sin \sigma \cos (\sigma +2 \tau )-\sin 2\sigma + 2 \sin 4\sigma \\&
    - 3 \sin 4\tau \bigg] + \frac{4 \theta_{13}}{\sqrt{2R}} \bigg[-3 (7 R+4) \sin 2(\sigma +\tau ) + 8
   \sqrt{R} \sin 2\sigma\\& - 16 \sqrt{R} \sin 4\sigma + 3 (5 R+4) \sin 2\tau \bigg]\, ,
\end{split}\\
\begin{split}\label{eq:AIOhigh_d}
    576\cdot A(m_{3}&=10^{-2}\ \text{eV}|\delta=3\frac{\pi}{2})=\widetilde{A}_{3} + 24\theta_{13}^2  \bigg[-\frac{3}{\sqrt{R}} \cos \sigma \sin (\sigma +2 \tau )+\sin 2\sigma -2 \sin 4\sigma\\&-3 \sin 4\tau \bigg] +\frac{4\theta_{13}}{\sqrt{2R}} \bigg[-3 (7 R+4) \cos 2 (\sigma +\tau )+8 \sqrt{R}(1+ \cos 2\sigma)\\& -16 \sqrt{R} \cos 4\sigma +15 R \cos 2\tau +12 \cos 2\tau \bigg]\, ,
\end{split}
\end{align}
}
with $\widetilde{A}_{3}=\bigg[16 \sin  2\sigma + 16 \sin 4\sigma +36 \sin 4\tau + \frac{6 (7 R+4)}{\sqrt{R}} \sin 2(\sigma +\tau )+\frac{3 (5 R+4)}{\sqrt{R}} \sin 2\tau \bigg]$\, .


\paragraph{Inverted ordering:} $m_{3}\sim10^{-4}\ \text{eV}$:
\small{
\begin{align}
\begin{split}\label{eq:AIOlow_gen}
    576 \cdot A(m_{3}&\sim10^{-4}\ \text{eV})\simeq \widetilde{A}_{4} + \frac{24\theta_{13}^2}{\sqrt{\eta }} \bigg[(2 \sin (\delta -\sigma )+\sin (\delta +\sigma )) \cos (\delta +\sigma +2 \tau )\\& 
    +\sqrt{\eta}(4 \cos 2\sigma -1) \sin2 (\delta +\sigma ) -3\sqrt{\eta} \sin 4\tau \bigg]\\&
   +32\theta_{13} \sin \sigma \bigg[\cos (\delta +\sigma )+2 \cos (\delta +3 \sigma ) +\frac{3}{\sqrt{2\eta}} \cos  (\delta +\sigma +2 \tau )\bigg]\, ,
\end{split}
\end{align}
\begin{align}
\begin{split}\label{eq:AIOlow_st}
    576 \cdot A(m_{3}&\sim10^{-4}\ \text{eV}|\sigma,\tau=0)\simeq 36 \theta_{13}^2 \sin 2\delta \left(2  +\frac{1}{\sqrt{\eta }}\right)\, ,
\end{split}\\
\begin{split}
    576 \cdot A(m_{3}&\sim10^{-4}\ \text{eV}|\delta=\pi)\simeq \widetilde{A}_{4} + 24 \theta_{13}^2 \bigg[-\frac{1}{\sqrt{\eta }}\sin \sigma  \cos (\sigma +2 \tau )-\sin 2\sigma\\&
    + 2 \sin 4\sigma  - 3 \sin 4\tau \bigg] -32\theta_{13}  \sin \sigma \bigg[2 \sqrt{\eta } \cos
   \sigma + 2 \cos 3\sigma + \frac{3}{ \sqrt{2\eta }} \cos (\sigma +2 \tau )\bigg]
\end{split}\\    
\begin{split}\label{eq:AIOlow_d}
    576 \cdot A(m_{3}&\sim10^{-4}\ \text{eV}|\delta=3\frac{\pi}{2})\simeq \widetilde{A}_{4} + 24 \theta_{13}^2  \bigg[-\frac{3}{\sqrt{\eta }} \cos \sigma \sin (\sigma +2 \tau )+\sin 2\sigma \\& -2 \sin 4\sigma -3 \sin 4\tau \bigg]
   +32 \theta_{13} \sin \sigma \bigg[\sin \sigma  +2 \sin 3 \sigma  + \frac{3}{\sqrt{2\eta}} \sin (\sigma +2 \tau )\bigg]\, ,
\end{split}
\end{align}
}
with $\widetilde{A}_{4}= \bigg[16\sin 2\sigma +16\sin 4\sigma + 36 \sin 4\tau  + \frac{6 (R+4)}{\sqrt{\eta }}  \sin 2(\sigma +\tau ) - \frac{3 (R-4)}{\sqrt{\eta }}  \sin 2\tau \bigg]$\, .


 \bibliographystyle{elsarticle-num} 
 \bibliography{cas-refs}
\end{document}